\documentclass[12pt,english]{article}
\usepackage[T1]{fontenc}
\usepackage[latin9]{inputenc}
\usepackage{geometry}
\usepackage{color}
\usepackage{babel}
\usepackage{amsmath}
\usepackage{amsthm}
\usepackage{amssymb}
\usepackage{graphicx}
\usepackage{setspace}
\usepackage[authoryear]{natbib}
\usepackage[unicode=true,pdfusetitle,
 bookmarks=true,bookmarksnumbered=false,bookmarksopen=false,
 breaklinks=false,pdfborder={0 0 1},backref=false,colorlinks=false]
 {hyperref}
\hypersetup{
 colorlinks=true,citecolor=blue, urlcolor=cyan, linkcolor=black}

\makeatletter
\theoremstyle{plain}
    \ifx\thechapter\undefined
      \newtheorem{assumption}{\protect\assumptionname}
    \else
      \newtheorem{assumption}{\protect\assumptionname}[chapter]
    \fi
\theoremstyle{definition}
    \ifx\thechapter\undefined
      \newtheorem{example}{\protect\examplename}
    \else
      \newtheorem{example}{\protect\examplename}[chapter]
    \fi
\theoremstyle{plain}
    \ifx\thechapter\undefined
	    \newtheorem{thm}{\protect\theoremname}
	  \else
      \newtheorem{thm}{\protect\theoremname}[chapter]
    \fi
\theoremstyle{plain}
    \ifx\thechapter\undefined
      \newtheorem{lyxalgorithm}{\protect\algorithmname}
    \else
      \newtheorem{lyxalgorithm}{\protect\algorithmname}[chapter]
    \fi

\usepackage{fontawesome}

\makeatother

\providecommand{\algorithmname}{Algorithm}
\providecommand{\assumptionname}{Assumption}
\providecommand{\examplename}{Example}
\providecommand{\theoremname}{Theorem}

\begin{document}
\date{}
\title{{\Huge{}\vskip -1em}Selecting invalid instruments to improve Mendelian
randomization with two-sample summary data\textsc{\huge{}}\thanks{We thank participants at the 2022 International Society for Clinical
Biostatistics conference, and Dipender Gill for helpful discussions.\protect \\
Ashish Patel (\protect\href{mailto:ashish.patel@mrc-bsu.cam.ac.uk}{ashish.patel@mrc-bsu.cam.ac.uk});
Francis J. DiTraglia (\protect\href{mailto:francis.ditraglia@economics.ox.ac.uk}{francis.ditraglia@economics.ox.ac.uk});
Verena Zuber (\protect\href{mailto:v.zuber@imperial.ac.uk}{v.zuber@imperial.ac.uk});
Stephen Burgess (\protect\href{http://sb452@medschl.cam.ac.uk}{sb452@medschl.cam.ac.uk})}}
\author{Ashish Patel{\small{} }\textsuperscript{{\small{}a}}{\small{},}
Francis J. DiTraglia{\small{} }\textsuperscript{{\small{}b}}, Verena
Zuber{\small{} }\textsuperscript{{\small{}c}}, \& Stephen Burgess{\small{}
}\textsuperscript{{\small{}a,d}}}
\maketitle
\begin{center}
{\small{}\vskip -2em}\textsuperscript{{\small{}a}}{\small{} MRC
Biostatistics Unit, University of Cambridge}\\
{\small{}}\textsuperscript{{\small{}b}}{\small{} Department of Economics,
University of Oxford}\\
{\small{}}\textsuperscript{{\small{}c}}{\small{} Department of Epidemiology
and Biostatistics, Imperial College London}\\
{\small{}}\textsuperscript{{\small{}d}}{\small{} Cardiovascular
Epidemiology Unit, University of Cambridge}{\small\par}
\par\end{center}
\begin{abstract}
Mendelian randomization (MR) is a widely-used method to estimate the
causal relationship between a risk factor and disease. A fundamental
part of any MR analysis is to choose appropriate genetic variants
as instrumental variables. Genome-wide association studies often reveal
that hundreds of genetic variants may be robustly associated with
a risk factor, but in some situations investigators may have greater
confidence in the instrument validity of only a smaller subset of
variants. Nevertheless, the use of additional instruments may be optimal
from the perspective of mean squared error even if they are slightly
invalid; a small bias in estimation may be a price worth paying for
a larger reduction in variance. For this purpose, we consider a method
for ``focused'' instrument selection whereby genetic variants are
selected to minimise the estimated asymptotic mean squared error of
causal effect estimates. In a setting of many weak and locally invalid
instruments, we propose a novel strategy to construct confidence intervals
for post-selection focused estimators that guards against the worst
case loss in asymptotic coverage. In empirical applications to: (i)
validate lipid drug targets; and (ii) investigate vitamin D effects
on a wide range of outcomes, our findings suggest that the optimal
selection of instruments does not involve only a small number of biologically-justified
instruments, but also many potentially invalid instruments. 
\end{abstract}
\newpage{}

\section{Introduction}

Mendelian randomization (MR) uses genetic variants as instrumental
variables to estimate the causal effect of a risk factor on an outcome
in the presence of unobserved confounding. By Mendel's second law,
genetic variants sort independently of other traits. Thus, genetic
variants, which are fixed at conception, can provide a source of exogenous
variation in a risk factor of interest, allowing analyses that are
less vulnerable to reverse causality and confounding \citep{Smith2003}. 

Large-scale consortia genome-wide association studies (meta-GWASs)
have identified large numbers of genetic variants that are robustly
associated with a wide range of traits. Due in part to privacy issues,
often only summary statistics of these genetic associations are made
publicly available. Since these results are easily accessible, MR
investigations increasingly rely on inferential methods that require
only two-sample summary data \citep{Burgess2015}. In such applications,
genetic variant associations with the risk factor are obtained from
a representative but non-overlapping sample used to measure genetic
variant associations with the outcome. 

As in usual instrumental variable analyses, identifying causal effects
through MR requires some key assumptions. For a genetic variant to
be a valid instrument, it must be associated with the risk factor;
$\beta_{X_{j}}\neq0$ in Figure 1 (relevance). Second, the variant
must be uncorrelated with unobserved confounders. Third, any effect
that the variant has on the outcome must be mediated by its effect
on the risk factor; $\tau_{j}=0$ in Figure 1 (exclusion).
\begin{center}
\includegraphics[width=8cm]{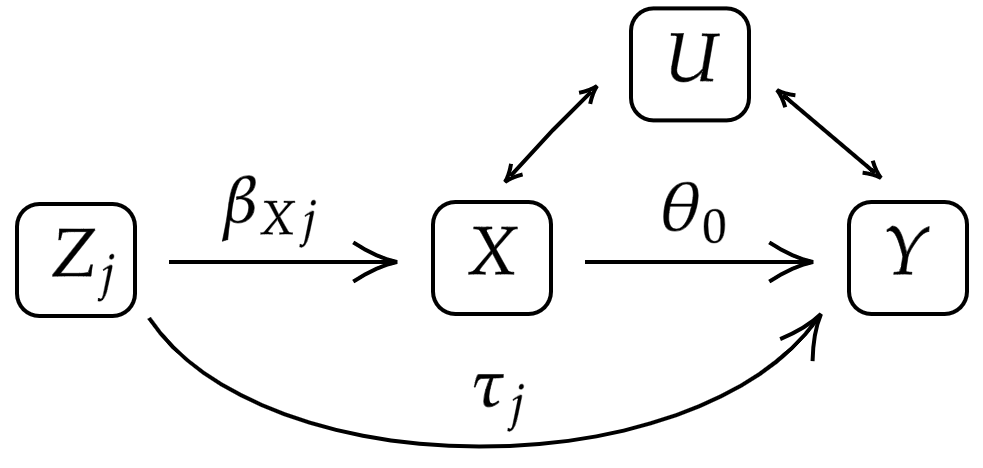}
\par\end{center}

\begin{center}
{\small{}Figure 1.}{\footnotesize{} The effect of genetic variant
$Z_{j}$ on the risk factor $X$ and outcome $Y$, where $U$ is an
unobserved confounder.}{\footnotesize\par}
\par\end{center}

Violations of the exclusion condition are common in MR studies \citep{Lawlor2008},
due in part to the widespread phenomenon of pleiotropy where a single
genetic variant may influence several traits \citep{Solovieff2013,Hemani2018}.
While the biological mechanism of certain genetic effects may be well
understood (for example, when using protein risk factors for drug
target validation; see \citealp{Schmidt2019}), it is generally difficult
to rule out the possibility that many genetic variants have a direct
effect on the outcome, and thus violate the exclusion restriction
\citep{Verbanck2018}. 

In some MR applications, investigators may have more confidence in
the validity of a particular subset of all candidate instruments.
In particular, the causal mechanisms linking specific genes to risk
factors may be known, which may better justify the use of genetic
variants from those genes as instruments. Examples of MR studies that
have prioritised the use of variants from biologically-plausible genes
include investigations into the effects of alcohol consumption \citep{Millwood2019},
C-reactive protein level \citep{Swerdlow2016}, smoking behaviour
\citep{Lassi2016}, vitamin D supplementation \citep{Mokry2015},
and perturbing drug targets \citep{Gill2021}. 

Along with biologically-justified instruments, investigators could
consider using additional variants that are plausibly valid instruments
in order to improve the precision of an analysis. Even if these additional
instruments are slightly invalid, their use may provide slightly biased
but more precise estimates that are optimal from the perspective of
mean squared error. 

Hence, to guide this instrument choice, we propose a method for ``focused''
instrument selection whereby genetic variants are selected to minimise
the estimated asymptotic mean squared error of causal effect estimates.
The strategy allows a \textsl{tiering} in the assumptions on instrument
validity, and prioritises the evidence suggested by biologically-justified
instruments. 

We work with the popular two-sample summary data design, and in a
setting of many weak and locally invalid instruments. In this local
misspecification setting, the collective direct instrument effects
on the outcome are decreasing with the sample size at a rate that
ensures a meaningful bias-variance trade off, and thus enables a mean
squared error comparison in finite samples. 

The theoretical contribution of our work has two elements. First,
we extend \citet{DiTraglia2016}'s results to allow for many weak
instruments. We consider an asymptotic framework in which the number
of instruments can grow at the same rate as the sample size, as long
as their collective effects on the risk factor are bounded \citep{Zhao2018}.
Second, we consider the problem of post-selection inference for focused
estimators which is particularly challenging because we do not have
consistent model selection; the uncertainty in model selection directly
impacts the asymptotic distribution of focused estimators. 

Given that focused use of additional instruments may lead to improved
estimation compared with using only a smaller set of biologically-justified
instruments, it is natural to consider whether a similar advantage
may hold for inference. One desirable property for confidence intervals
is that they are \textsl{uniformly valid}; that is, they achieve nominal
coverage asymptotically over the space of potential direct instrument
effects on the outcome. Unfortunately, the impossibility results discussed
by \citet{Leeb2005} suggest we cannot construct uniformly valid confidence
intervals for focused estimators which will exactly achieve nominal
coverage, and therefore such intervals will generally be conservative.
For example, in simulation we find that the ``2-step'' uniformly
valid confidence intervals of \citet{DiTraglia2016} can be over 30\%
longer than standard confidence intervals based on using only a core
set of valid instruments.

In the same way that focused estimation is willing to trade a small
bias in estimation for a larger reduction in variance, we develop
a strategy for constructing ``focused'' confidence intervals that
accepts a small loss in coverage probability over a subspace of direct
instrument effects on the outcome in exchange for shorter confidence
intervals. Our focused confidence intervals account for uncertainty
in model selection, and they achieve nominal coverage for plausible
values of direct instrument effects, while providing a statistical
guarantee on the worst case size distortion away from those values. 

Compared with using only a core set of biologically-justified instruments,
our simulation evidence illustrates how focused estimators and confidence
intervals may provide improved estimation and inference when additional
instruments are valid or slightly invalid, while \textsl{buying insurance}
against poor performance when additional instruments are very invalid. 

The utility of our methods is demonstrated in two empirical applications.
The first considers genetic validation of lipid drug targets when
investigators are uncertain about defining the width of a \textsl{cis}-gene
window from which to select instruments. The second application investigates
the genetically predicted effect of vitamin D supplementation on a
range of outcomes when investigators want to prioritise the use of
genetic variants from specific genes through biological considerations. 

We use the following notation and abbreviations: $\overset{P}{\to}$
`converges in probability to'; $\overset{D}{\to}$ `converges in distribution
to'; $\overset{a}{\sim}$ `is asymptotically distributed as'. For
any sequences $a_{n}$ and $b_{n}$, if $a_{n}=O(b_{n})$, then there
exists a positive constant $M$ and a positive integer $N$ such that
for all $n\geq N$, $b_{n}>0$ and $\vert a_{n}\vert\leq Mb_{n}$.
If $a_{n}=o(b_{n})$, then $\vert a_{n}\vert/b_{n}\to0$ as $n\to\infty$.
Also, if $a_{n}=\Theta(b_{n})$, then there exist positive constants
$M_{1}$ and $M_{2}$, $M_{1}\leq M_{2}<\infty$, and a positive integer
$N$ such that $M_{1}b_{n}\leq a_{n}\leq M_{2}b_{n}$ for all $n\geq N$.
The proofs of theoretical results are given in Appendix, and R code
to perform our empirical investigation is available on GitHub at github.com/ash-res/focused-MR/. 

\section{Model and assumptions}

\subsection{Two-sample summary data}

We first outline our assumptions on genetic association summary data,
which are motivated through a simple linear model with invalid instruments
and homoscedastic errors.

Let $Z=(Z_{1},...,Z_{p})^{\prime}$ denote a $p$-vector of uncorrelated
genetic variants. The parameter of interest is the causal effect $\theta_{0}$
of the risk factor $X$ on the outcome $Y$, which is described by
\begin{eqnarray}
Y & = & \omega_{Y}+X\theta_{0}+Z^{\prime}\tau+U_{Y}\\
X & = & \omega_{X}+Z^{\prime}\beta_{X}+U_{X},
\end{eqnarray}
where $E[U_{Y}\vert Z]=0$, $E[U_{X}\vert Z]=0$, $E[U_{Y}^{2}\vert Z]=\sigma_{U_{Y}}^{2}$,
$E[U_{X}^{2}\vert Z]=\sigma_{U_{X}}^{2}$, $\beta_{X}=(\beta_{X_{1}},\ldots,\beta_{X_{p}})^{\prime}$,
and $(\omega_{X},\omega_{Y},\beta_{X},\tau,\theta_{0})$ are unknown
parameters. If $\tau$ is non-zero, then at least one genetic variant
fails the exclusion restriction and directly affects the outcome.

Substituting $(2)$ into $(1)$, we have $Y=(\omega_{Y}+\omega_{X}\theta_{0})+Z^{\prime}(\beta_{X}\theta_{0}+\tau)+(U_{Y}+U_{X}\theta_{0})$.
Thus, $Cov(Z,Y)=Var(Z)(\beta_{X}\theta_{0}+\tau)$, which leads to
a model 
\begin{equation}
\beta_{Y}=\beta_{X}\theta_{0}+\tau,
\end{equation}
where $\beta_{Y}=(\beta_{Y_{1}},\ldots,\beta_{Y_{p}})^{\prime}$ is
the $p$-vector of coefficients from a population regression of $Y$
on $Z$. We aim to estimate the model in $(3)$ using two-sample summary
data on genetic associations.
\begin{assumption}[two-sample summary data]
 For each variant $j$, we observe genetic associations $\hat{\beta}_{X_{j}}$
and $\hat{\beta}_{Y_{j}}$, which satisfy $\hat{\beta}_{X_{j}}\sim N(\beta_{X_{j}},\sigma_{X_{j}}^{2})$
and $\hat{\beta}_{Y_{j}}\sim N(\beta_{Y_{j}},\sigma_{Y_{j}}^{2})$,
where $\sigma_{X_{j}}^{2}=\Theta(1/n_{X})$ and $\sigma_{Y_{j}}^{2}=\Theta(1/n_{Y})$
are assumed to be known. Moreover, the set of $2p$ genetic associations
$\{\hat{\beta}_{X_{j}},\hat{\beta}_{Y_{j}}\}_{j=1}^{p}$ are mutually
uncorrelated, and $n_{X}/n_{Y}\to c$, as $n:=(n_{X},n_{Y})\to\infty$
for some constant $0<c<\infty$. 
\end{assumption}
Assumption 1 is taken from \citet{Zhao2018} and states a normal approximation
of estimated genetic associations which is typically justified by
large random sampling expected in genetic association studies. Specifically,
for each variant $j$, we have access to estimates and standard errors
from univariable $Z_{j}$ on $X$ linear regressions from an $n_{X}$-sized
sample, and from a non-overlapping $n_{Y}$-sized sample, we observe
measured associations from univariable $Z_{j}$ on $Y$ linear regressions.
Both random samples are drawn from the joint distribution of $(Y,X,Z)$. 

The assumption that the population standard deviations $\{\sigma_{X_{j}},\sigma_{Y_{j}}\}_{j=1}^{p}$
are known is common in MR, and a formal justification for this is
given by \citet{Ye2020}. To simplify notation, the dependence of
the standard errors on the sample size is not made explicit, but they
are assumed to decrease at the usual parametric rate. 

\subsection{Core instruments}

From $(3)$, the parameter of interest $\theta_{0}$ is not identified
unless there are some restrictions on $\tau=(\tau_{1},\ldots,\tau_{p})^{\prime}$.
When genetic variants from several gene regions are used to instrument
the risk factor, a popular identification strategy is to assume that
$\tau$ is a mean zero random effect; see, for example, \citet{Zhao2018}.
Another commonly-used assumption is that most genetic variants are
valid instruments; $\tau_{j}=0$ for $j\in S_{M}$, where $S_{M}$
is some unknown set of variants such that $p^{-1}\vert S_{M}\vert>0.5$.
In this setting the median-based estimators of \citet{Bowden2016}
are consistent. Based on variation of these assumptions, many summary
data MR methods have proposed ways to obtain unbiased estimates of
$\theta_{0}$; a recent review is given in \citet{Sanderson2022}. 

In this work we do not focus on a novel identification strategy, but
instead only the simple instrument selection choice for investigators
when they believe a \textsl{core} set of genetic variants $S_{0}$
consists of only valid instruments, but they are less confident on
the validity of any additional candidate instruments. 
\begin{assumption}[core instruments]
 For some known set of instruments $S_{0}$ such that $1\leq\vert S_{0}\vert<p$,
and where $\vert S_{0}\vert$ grows proportionately with $p$, we
have $\tau_{j}=0$ for all $j\in S_{0}$. 
\end{assumption}
Assumption 2 allows us to measure the bias which may result from the
inclusion of additional instruments. In some MR studies, there is
good reason for believing that a smaller subset of all available genetic
variants are more likely to be valid instruments. We mention two examples
that we explore in more detail in Section 6. 
\begin{example}[choosing \textsl{cis} windows in drug target MR]
 In drug target MR studies, only those genetic variants from a single
gene region that encodes the protein target of a drug are used to
instrument the risk factor. Such studies have gained popularity for
providing supporting genetic evidence to validate drug targets and
to study side effects \citep{Gill2021}. A key decision in drug target
MR is to choose the width of a ``\textsl{cis w}indow'' which dictates
a gene region from which to select instruments. Tools such as GeneCards
\citep{Stelzer2016} offer practical guidance on defining appropriate
gene regions, but since there are often relatively few uncorrelated
genetic signals from narrow \textsl{cis w}indows, researchers often
resort to widening the \textsl{cis w}indow in order to boost power
through the use of a larger number of instruments. This leaves the
possibility of a type of publication bias where researchers may report
findings only from a \textsl{cis w}indow which gives their preferred
result. We may be less confident on the instrument validity of additional
genetic variants that are included only from widening a \textsl{cis
w}indow. 
\begin{example}[estimating vitamin D effects]
 GWASs have identified strong genetic associations with vitamin D
in biologically plausible genes; \textsl{GC}, \textsl{DHCR7}, \textsl{CYP2R1},
and \textsl{CYP24A1}. Each of these genes is known influence vitamin
D level through different mechanisms. In order to investigate the
effect of vitamin D supplementation on a range of traits and diseases,
MR studies have often used genetic variants located in neighborhoods
of those genes to instrument vitamin D as they are considered more
likely to satisfy the exclusion restriction than other genome-wide
significant variants \citep{Mokry2015,Revez2020}. Genetic variants
from other gene regions may be strongly associated with vitamin D,
but they may be more likely to have direct effects on the outcome
through their effects on traits other than vitamin D. 
\end{example}
\end{example}

\subsection{Many weak variant associations}

In typical applications, we may expect many genetic variants to have
weak effects on the risk factor. This can cause difficulties for identifying
and estimating the causal effect. We study a setting of many weak
instruments where the number of genetic variants is permitted to grow
at the same rate as the sample size, but their collective explanatory
power is bounded. 
\begin{assumption}[many weak instruments]
 Let $\beta_{X,0}$ denote a $\vert S_{0}\vert$-vector with its
elements given by $\beta_{X_{j}}$ for $j\in S_{0}$. Then, $\Vert\beta_{X}\Vert_{2}=O(1)$,
$\Vert\beta_{X}\Vert_{3}\big/\Vert\beta_{X,0}\Vert_{2}\to0$, and
$p\big/n\Vert\beta_{X,0}\Vert_{2}^{2}=O(1)$ as $n,p\to\infty$.
\end{assumption}
Given $(3)$ and Assumptions 2 and 3, the causal effect is point identified.
Assumption 3 is similar to \citet{Zhao2018}, and it implies that
all variants are relevant instruments, but the explanatory power of
any individual variant is decreasing as $p\to\infty$. Moreover, the
skewness of $\beta_{X}$ is restricted which rules out very sparse
variant effects settings. Finally, the rate restriction $p\big/n\Vert\beta_{X,0}\Vert_{2}^{2}=O(1)$
as $n,p\to\infty$ ensures asymptotic normality in estimation, and
the condition is plausible given the large sample sizes of typical
GWASs. 

\subsection{Additional locally invalid instruments}

To meaningfully study a bias-variance trade off from including additional
instruments we need to ensure that the bias and variance terms have
the same order of magnitude. Following \citet{DiTraglia2016}'s framework
of \textsl{locally invalid} instruments, we work with local misspecification
in which the direct effects $\tau$ are collectively ``local-to-zero'',
and decrease at the same rate as the sampling errors with $n$. This
is not a substantive biological assumption, but rather a technical
device that allows us to study the instrument selection problem from
a mean squared error perspective. 
\begin{assumption}[locally invalid instruments]
 $\Vert\tau\Vert_{2}^{2}=O(1\big/n)$ as $n,p\to\infty$.
\end{assumption}
Under the rate restriction in Assumption 4, we can consistently estimate
$\theta_{0}$ using any set of instruments, but making valid inferences
using invalid instruments will require us to account for an asymptotic
bias. The rate restriction is slightly different to \citet{DiTraglia2016}
in that we limit only the collective direct effects over all variants. 

Assumption 4 is otherwise quite general. For example, the direct effects
$\tau$ may be correlated with $\beta_{X}$, and thus may violate
the so-called InSIDE (Instrument Strength Independent of Direct Effect)
assumption which some existing MR methods that make use of invalid
instruments rely on \citep{Bowden2015}. 

\section{Focused instrument selection and estimation}

Using only the core set of instruments $S_{0}$ should lead to asymptotically
unbiased estimation of $\theta_{0}$, and this forms a basis from
which to measure the potential bias from including additional instruments.
For ease of exposition we focus on the simple choice between using
either: (i) the ``Core'' estimator that chooses only the set of
core instruments $S_{0}$; or (ii) the ``Full'' estimator that uses
full set of $p$ instruments. The inclusion of $S$, the set of the
additional $p-\vert V\vert$ genetic variants, may result in improved
estimation if there is a large reduction in variance, and at most
only a small increase in bias. Thus we use a ``focused'' instrument
selection strategy \citep{DiTraglia2016} where the Full estimator
is selected only if it has a lower estimated asymptotic mean squared
error than the Core estimator.

The results presented here can be easily extended for more fine-tuned
instrument selection at the expense of extra notation; instead of
simply choosing between the Core and Full estimator, we could also
select from subsets of the additional instruments. Such subsets could
be data-driven, chosen on biological reasons, and have overlapping
variants. For example, in our empirical application to study vitamin
D effects, we chose from all possible subsets of 3 partitions of the
additional instruments, where the partitions were formed by k-means
clustering on the Wald ratio estimate of each variant, $\hat{\beta}_{Y_{j}}\big/\hat{\beta}_{X_{j}}$,
$j\in S$. 

We consider limited information maximum likelihood (LIML) estimation
of $\theta_{0}$. The Core and Full estimators, $\hat{\theta}_{C}$
and $\hat{\theta}_{F}$, are given by
\[
\hat{\theta}_{C}=\arg\min_{\theta}\sum_{j\in S_{0}}\frac{(\hat{\beta}_{Y_{j}}-\hat{\beta}_{X_{j}}\theta)^{2}}{\sigma_{Y_{j}}^{2}+\sigma_{X_{j}}^{2}\theta^{2}}\,\,\,\,\,\,\text{and}\,\,\,\,\,\,\hat{\theta}_{F}=\arg\min_{\theta}\sum_{j\in S_{0}\cup S}\frac{(\hat{\beta}_{Y_{j}}-\hat{\beta}_{X_{j}}\theta)^{2}}{\sigma_{Y_{j}}^{2}+\sigma_{X_{j}}^{2}\theta^{2}}.
\]
Under Assumptions 1-3, Theorem 3.1 of \citet{Zhao2018} shows that
the asymptotic distribution of $\hat{\theta}_{C}$ is 
\[
\Delta_{C}^{-1/2}(\hat{\theta}_{C}-\theta_{0})\overset{D}{\to}N(0,1),
\]
as $n,p\to\infty$, where $\Delta_{C}=\eta_{C}^{-2}(\eta_{C}+\varsigma_{C})$,
$\eta_{C}=\sum_{j\in S_{0}}\Omega_{j}^{-1}\beta_{X_{j}}^{2}$ $\varsigma_{C}=\sum_{j\in S_{0}}\Omega_{j}^{-2}\sigma_{X_{j}}^{2}\sigma_{Y_{j}}^{2}$,
and $\Omega_{j}=\sigma_{Y_{j}}^{2}+\theta_{0}^{2}\sigma_{X_{j}}^{2}$.
Using similar arguments, we can derive the asymptotic distribution
of the Full estimator. 
\begin{thm}[Full estimator]
 Under Assumptions 1-4, and the Full estimator is consistent, and
its asymptotic distribution is
\[
\Delta_{F}^{-1/2}(\hat{\theta}_{F}-\theta_{0}-b)\overset{D}{\to}N(0,1),
\]
as $n,p\to\infty$, where $\Delta_{F}=(\eta_{C}+\eta_{S})^{-2}(\eta_{C}+\eta_{S}+\varsigma_{C}+\varsigma_{S})$,
$b=(\eta_{C}+\eta_{S})^{-1}b_{S}$, $b_{S}=\sum_{j\in S}\Omega_{j}^{-1}\beta_{X_{j}}\tau_{j}$,
$\eta_{S}=\sum_{j\in S}\Omega_{j}^{-1}\beta_{X_{j}}^{2}$, and $\varsigma_{S}=\sum_{j\in S}\Omega_{j}^{-2}\sigma_{X_{j}}^{2}\sigma_{Y_{j}}^{2}$. 
\end{thm}
The variance terms $\eta_{C}$ and $\eta_{S}$ are of order $O(n\Vert\beta_{X}\Vert_{2}^{2})$,
and the terms $\varsigma_{C}$ and $\varsigma_{S}$ are of order $O(p)$.
Therefore, $(\eta_{C}+\eta_{S})^{-1}$ would be the asymptotic variance
of $\hat{\theta}_{F}$ in a fixed $p$ setting with strong instruments,
and $(\eta_{C}+\eta_{S})^{-2}(\varsigma_{C}+\varsigma_{S})$ is an
additional variance component to account for the extra uncertainty
due to many weak instruments \citep{Zhao2018}. This is particularly
important for our instrument selection problem since under-estimated
variances based on `fixed $p$' asymptotics could cause a mean squared
error-based selection criteria to falsely recommend the inclusion
of additional instruments. These two variance components are of the
same order of magnitude when $p\big/n\Vert\beta_{X}\Vert_{2}^{2}=\Theta(1)$.

As discussed by \citet{Newey2009a}, despite the knife-edge condition
required to balance these variance components, it may be advisable
to use the weak instrument variance correction $(\eta_{C}+\eta_{S})^{-2}(\varsigma_{C}+\varsigma_{S})$
in general scenarios when $n$ is considerably larger than $p$, and
when instruments are strong. For example, the simulation study of
\citet[pp. 457-9]{Davies2015}, which mimicked an MR design, showed
that standard errors that did not correct for many weak instrument
effects led to inflated type I error rates even for the case where
$n=3000$ and $p=9$. 

We also note that compared with the model of \citet{Zhao2018}, here
we consider $\tau$ to be fixed effect rather than a random variable;
this direct variant effect on the outcome induces a bias in estimation
rather than an increase in variance. The asymptotic variance of $\hat{\theta}_{F}$
is of order $O(1\big/n\Vert\beta_{X}\Vert_{2}^{2})+O(p\big/n^{2}\Vert\beta_{X}\Vert_{2}^{4})$,
and its asymptotic bias is of order $O(1\big/\sqrt{n}\Vert\beta_{X}\Vert_{2})$.
Thus, there is a meaningful bias-variance trade off if $p\big/n\Vert\beta_{X}\Vert_{2}^{2}=O(1)$
since the square of the asymptotic bias is of the same order of magnitude
as the asymptotic variance. 

To carry out focused instrument selection, we need to estimate and
compare the asymptotic mean squared error (AMSE) of $\hat{\theta}_{C}$
and $\hat{\theta}_{F}$. The Core estimator is asymptotically unbiased,
and it is straightforward to consistently estimate its asymptotic
variance $\Delta_{C}$. Under local misspecification, the asymptotic
bias $b$ of the Full estimator cannot be consistently estimated.
However, we can use $\hat{\theta}_{C}$ to construct an asymptotically
unbiased estimate of $b$. 

Let $\hat{\eta}_{C}=\sum_{j\in S_{0}}\hat{\Omega}_{j}^{-1}(\hat{\beta}_{X_{j}}^{2}-\sigma_{X_{j}}^{2})$,
$\hat{\eta}_{S}=\sum_{j\in S}\hat{\Omega}_{j}^{-1}(\hat{\beta}_{X_{j}}^{2}-\sigma_{X_{j}}^{2})$,
and $\hat{\Omega}_{j}=\sigma_{Y_{j}}^{2}+\hat{\theta}_{C}^{2}\sigma_{X_{j}}^{2}$.
Then, our estimator of $b$ is $\hat{b}=(\hat{\eta}_{C}+\hat{\eta}_{S})^{-1}\hat{b}_{S}$,
where $\hat{b}_{S}=\sum_{j\in S}\hat{\Omega}_{j}^{-1}\hat{\beta}_{X_{j}}\hat{\beta}_{Y_{j}}-\hat{\theta}_{C}\sum_{j\in S}\hat{\Omega}_{j}^{-1}(\hat{\beta}_{X_{j}}^{2}-\sigma_{X_{j}}^{2})$. 
\begin{thm}[Asymptotic bias estimator]
 Under Assumptions 1-4, the asymptotic distribution of $\hat{b}$
is 
\[
\Delta_{B}^{-1/2}(\hat{b}-b)\overset{D}{\to}N(0,1),
\]
as $n,p\to\infty$, where $\Delta_{B}=(\eta_{C}+\eta_{S})^{-2}[\eta_{S}+\varsigma_{S}+\xi_{S}+\eta_{C}^{-2}\eta_{S}^{2}(\eta_{C}+\varsigma_{C})]$
and $\xi_{S}=2\theta_{0}^{2}\Sigma_{j\in S}\Omega_{j}^{-2}\sigma_{X_{j}}^{4}$.
\end{thm}
Although we cannot consistently estimate $b$, Theorem 2 shows that
$\hat{b}$ is an asymptotically unbiased estimator. Similar to before,
we can write the asymptotic variance $\Delta_{B}$ as the sum of two
terms: $(\eta_{C}+\eta_{S})^{-2}\eta_{S}(1+\eta_{C}^{-1}\eta_{S})$
and $(\eta_{C}+\eta_{S})^{-2}(\varsigma_{S}+\xi_{S}+\eta_{C}^{-2}\eta_{S}^{2}\varsigma_{C})$.
The first term is the asymptotic variance of $\hat{b}$ in a fixed
$p$ setting with strong instruments, and therefore the second term
represents the extra uncertainty in estimation due to many weak instruments. 

In order to estimate the AMSE of $\hat{\theta}_{C}$ and $\hat{\theta}_{F}$,
we need to construct consistent estimators $\hat{\Delta}_{C}=\hat{\eta}_{C}^{-2}(\hat{\eta}_{C}+\hat{\varsigma}_{C})$
for $\Delta_{C}$, $\hat{\Delta}_{F}=(\hat{\eta}_{C}+\hat{\eta}_{S})^{-2}(\hat{\eta}_{C}+\hat{\eta}_{S}+\hat{\varsigma}_{C}+\hat{\varsigma}_{S})$
for $\Delta_{F}$, and $\hat{\Delta}_{B}=(\hat{\eta}_{C}+\hat{\eta}_{S})^{-2}[\hat{\eta}_{S}+\hat{\varsigma}_{S}+\hat{\xi}_{S}+\hat{\eta}_{C}^{-2}\hat{\eta}_{S}^{2}(\hat{\eta}_{C}+\hat{\varsigma}_{C})]$
for $\Delta_{B}$ where $\hat{\varsigma}_{C}=\sum_{j\in S_{0}}\hat{\Omega}_{j}^{-2}\sigma_{X_{j}}^{2}\sigma_{Y_{j}}^{2}$,
$\hat{\varsigma}_{S}=\sum_{j\in S}\hat{\Omega}_{j}^{-2}\sigma_{X_{j}}^{2}\sigma_{Y_{j}}^{2}$,
and $\hat{\xi}_{S}=2\hat{\theta}_{C}^{2}\sum_{j\in S}\hat{\Omega}_{j}^{-2}\sigma_{X_{j}}^{4}$. 

Since $\hat{\theta}_{C}$ is asymptotically unbiased, a consistent
estimator for its AMSE is $\hat{\Delta}_{C}$. Whereas for $\hat{\theta}_{F}$,
an asymptotically unbiased estimator of its AMSE is $(\hat{b}^{2}-\hat{\Delta}_{B})+\hat{\Delta}_{F}$.
Following \citet{DiTraglia2016}, since the square of the asymptotic
bias cannot be negative, we use $\max(0,\hat{b}^{2}-\hat{\Delta}_{B})$
instead of $\hat{b}^{2}-\hat{\Delta}_{B}$ when estimating the AMSE
of $\hat{\theta}_{F}$. 

Define $\hat{W}=\max(\hat{b}^{2}-\hat{\Delta}_{B},0)+\hat{\Delta}_{F}-\hat{\Delta}_{C}$
as the estimated AMSE of $\hat{\theta}_{F}$ minus the estimated AMSE
of $\hat{\theta}_{C}$. Therefore, the selection event that we select
the Full estimator is given by $\{\hat{W}\leq0\}$. The ``Focused''
estimator is then given by 
\[
\hat{\theta}=I\{\hat{W}\leq0\}\hat{\theta}_{F}+(1-I\{\hat{W}\leq0\})\hat{\theta}_{C}.
\]
Since both $\hat{\theta}_{F}$ and $\hat{\theta}_{C}$ are consistent
estimators of $\theta_{0}$, so is the Focused estimator $\hat{\theta}$. 

\section{Post-selection inference}

In this section we discuss the problem of constructing confidence
intervals for the Focused estimator $\hat{\theta}$, which is non-standard
because model selection is based on an AMSE criterion that is not
consistently estimated. 

We start by deriving the asymptotic distribution of $\hat{\theta}$,
and discuss why naive confidence intervals that ignore sampling uncertainty
in instrument selection are likely to perform poorly. We then note
the relative merits of two related inference procedures proposed in
\citet{DiTraglia2016}, before introducing a new ``Focused'' approach
that combines their strengths. Just as the Focused estimator aims
to achieve a good balance between bias and variance, Focused intervals
aim to achieve a good balance between the length of confidence intervals
and the potential worst case asymptotic coverage over the likely space
of direct instrument effects on the outcome. 

\subsection{Asymptotic distribution of the Focused estimator}

A feature of the local misspecification framework is that the sampling
uncertainty from asymptotic bias estimation directly affects the asymptotic
distribution of the post-selection Focused estimator $\hat{\theta}$.
In particular, there is non-ignorable uncertainty in model selection:
even if the use of only the core instruments is optimal from an AMSE
perspective, the uncertainty in asymptotic bias estimation can cause
the Focused estimator to erroneously select the full set of instruments. 

In contrast, under consistent model selection, the Focused estimator
would choose either $\hat{\theta}_{C}$ or $\hat{\theta}_{F}$ with
probability approaching 1 as $n,p\to\infty$. While this would seem
to simplify the task of inference, it would still not be possible
to consistently estimate the distribution of post-selection estimators
uniformly over the space of direct effects $\tau$ \citep[Section 2.3, pp.  38--40]{Leeb2005}.
Moreover, consistent model selection does not suit our goal of improved
estimation in terms of low risk, since the worst case risk of post-selection
estimators would be unbounded \citep{Leeb2008}. 
\begin{thm}[Focused estimator]
 Under Assumptions 1-4, the Focused estimator is consistent, and
is asymptotically distributed $\hat{\theta}\overset{a}{\sim}\theta_{0}+\Lambda(b)$
as $n,p\to\infty$ where $\Lambda(b)=I\{W(b)\leq0\}(b+{\cal K}_{F})+(1-I\{W(b)\leq0\}){\cal K}_{C}$,
$W(b)=\max((b+{\cal K}_{b})^{2}-\Delta_{B},0)+\Delta_{F}-\Delta_{C}$,
and 
\[
{\cal K:=}\begin{bmatrix}{\cal K}_{F}\\
{\cal K}_{C}\\
{\cal K}_{b}
\end{bmatrix}\sim N\left(\begin{bmatrix}0\\
0\\
0
\end{bmatrix},\Delta:=\begin{bmatrix}\Delta_{F} & \Delta_{E} & \Delta_{D}\\
\Delta_{E} & \Delta_{C} & \Delta_{A}\\
\Delta_{D} & \Delta_{A} & \Delta_{B}
\end{bmatrix}\right),
\]
where $\Delta_{E}=\eta_{C}^{-1}(\eta_{C}+\eta_{S})^{-1}(\eta_{C}+\varsigma_{C})$,
$\Delta_{A}=-\Delta_{E}\eta_{C}^{-1}\eta_{S}$, and $\Delta_{D}=\eta_{C}^{-1}(\eta_{C}+\eta_{S})^{-2}\eta_{C}(\eta_{S}+\varsigma_{S})-\Delta_{E}(\eta_{C}+\eta_{S})^{-1}\eta_{S}$. 
\end{thm}
Theorem 3 indicates that the asymptotic distribution of $\hat{\theta}$
is a weighted average of the asymptotic distributions of $\hat{\theta}_{C}$
and $\hat{\theta}_{F}$, where the weights are random even as $n,p\to\infty$.
As a result, ``naive'' confidence intervals for $\hat{\theta}$
that ignore sampling uncertainty in instrument selection should not
be reported. Such intervals can perform extremely poor in practice,
with coverage arbitrarily far below their nominal level. 

Under the condition $p\big/n\Vert\beta_{X}\Vert_{2}^{2}=O(1)$ as
$n,p\to\infty$ from Assumption 3, the variance components $\Delta$
can be consistently estimated, and therefore to simplify notation
we henceforth assume that $\Delta$ is known. 

\subsection{Focused confidence intervals with coverage constraints}

There are two related problems that a valid inference procedure must
solve. First, confidence intervals need to be widened to account for
the model selection uncertainty introduced by focused instrument selection.
Second, confidence intervals need to be re-centered if the focused
estimator is asymptotically biased. 

If the true asymptotic bias component $b$ was known, inference would
be straightforward: Theorem 3 could be used directly to simulate the
distribution of $\Lambda(b)$. However, $b$ cannot be consistently
estimated in our locally invalid instruments framework. For this setting,
\citet{DiTraglia2016} proposes two feasible inference procedures
that consider the distribution of $\Lambda(b)$ at values of $b$
that are \textsl{plausible} given the observed data. 

The ``1-step'' interval is based on the distribution of $\Lambda(b)$
evaluated at $b=\hat{b}$, and it effectively assumes that the true
value of $b$ is equal to its asymptotically unbiased estimator $\hat{b}$.
This is intuitive since $\hat{b}$ is in some sense the \textsl{most}
plausible value of $b$ given the data. It follows that if $\hat{b}$
is close to the true value of $b$, then the 1-step interval will
have coverage that is close to its nominal level. Moreover, when the
direct effects $\tau$ are small, simulation evidence from \citet{DiTraglia2016}
suggests that the 1-step interval has competitive coverage and can
be shorter in length than the ``Core'' interval, which is the standard
confidence interval of $\hat{\theta}_{C}$ that uses only the core
instruments. More generally, however, the 1-step interval comes with
no theoretical guarantees; it may substantially under-cover, although
its performance is much better than that of a naive interval that
ignores instrument selection. 

The reason why the 1-step interval may under-cover is that it fails
to account for the uncertainty in the estimate $\hat{b}$ of $b$,
thus potentially leading to intervals that are too short and centered
incorrectly. The ``2-step'' interval allows for such uncertainty
by first constructing a confidence region $\varphi$ for $b$. It
then simulates the distribution of $\Lambda(b')$ at every value $b'$
in $\varphi$, constructing a collection of confidence intervals each
based on the assumption that the true value of $b$ is $b'$. To obtain
a uniform coverage guarantee, the 2-step interval takes the outer
envelope of \textsl{all} of the resulting intervals. This makes the
2-step interval extremely conservative: in general there is \textsl{no}
value of $b$ for which the actual coverage equals the nominal coverage,
and hence it will always over-cover. Our simulation evidence suggests
that this over-coverage problem makes the 2-step intervals too wide
to be useful in practice.

We consider a way forward for improved inference with ``Focused''
intervals. These intervals aim to combine the strengths of the 1-step
and 2-step intervals while avoiding their drawbacks. Like the 1-step
interval, it is constructed by simulating $\Lambda(b)$ at a \textsl{single}
value of $b$ rather than taking an outer envelope over many values.
This means that it can yield shorter confidence intervals. Like the
2-step interval, however, it comes with theoretical guarantees. The
key is to choose an appropriate value of $b$. 

The Focused interval considers only values of $b$ that are contained
in a $(1-\alpha_{1})\times100\%$ confidence interval called $\varphi$.
This is the same confidence interval for $b$ used in the 2-step interval
approach. For some values of $b$ in $\varphi$ the distribution of
$\Lambda(b)$ will be highly dispersed. Suppose that $b'$ is such
a value, so that a $(1-\alpha_{2})\times100\%$ confidence interval
$CI(b')$ for $\theta_{0}$ computed under the assumption that $b=b'$
will be relatively wide. By construction, $CI(b')$ will achieve nominal
coverage probability $1-\alpha_{2}$ when $b=b'$. The key insight
is as follows: if $b''$ is a value in $\varphi$ for which $\Lambda(b'')$
is relatively less dispersed, then $CI(b')$ may also be a \textsl{nearly}
valid confidence interval for $\theta_{0}$ when $b=b''$. Using this
idea, the construction of the Focused interval proceeds as follows,
based on a user-specified tolerance $\gamma$ and nominal coverage
$1-\alpha$.
\begin{lyxalgorithm}[Focused interval with a minimum coverage constraint]
~\\
\textbf{1.} Construct a $(1-\alpha_{1})\times100\%$ confidence interval
$\varphi$ for $b$ using Theorem 2. \\
\textbf{2.} For each $b'\in\varphi$, calculate a collection of $(1-\alpha_{2})\times100\%$
intervals $[a_{l}(b'),a_{u}(b')]$ for $\Lambda(b')$ using Theorem
3, each under the assumption that $b=b'$.\\
\textbf{3.} Calculate $\bar{\varphi}=\{b'\in\varphi:P(a_{l}(b')\leq\Lambda(b'')\leq a_{u}(b'))\geq1-\alpha_{2}-\gamma,\,\text{for all}\,b''\in\varphi\}$.\\
\textbf{4.} Find the value of $b'\in\bar{\varphi}$ that yields the
shortest $(1-\alpha_{2})\times100\%$ confidence interval for $\Lambda(b')$.
Call this value $b^{\star}$. A confidence interval for $\theta_{0}$
is then $[a_{l}(b^{\star}),a_{u}(b^{\star})]$.\\
\textbf{5.} Notice that $b^{\star}$ depends on $\gamma$, $\alpha_{1}$,
and $\alpha_{2}$. Repeat steps 1-4 for a range of choices of $\alpha_{1}$
subject to the constraint $\alpha_{1}+\alpha_{2}=\alpha$. Choose
the value of $\alpha_{1}$ that yields the shortest interval for $\theta_{0}$. 
\end{lyxalgorithm}
Like the 1-step interval, the Focused interval is \textsl{always}
shorter than the 2-step interval because it is one of the intervals
contained in the outer envelope that forms the 2-step interval. Unlike
the 1-step interval, its asymptotic coverage probability can never
fall below $1-\alpha-\gamma$. 
\begin{thm}[Worst case asymptotic coverage of Focused intervals]
 Under Assumptions 1-4, the Focused interval defined in Algorithm
1 has asymptotic coverage probability no less than $1-\alpha-\gamma$
as $n,p\to\infty$. 
\end{thm}
The Focused interval is designed to achieve nominal coverage $1-\alpha$
at a plausible value of the asymptotic bias, while also controlling
the \textsl{worst case} asymptotic coverage according to a maximum
allowable size distortion $\gamma$. The choice of $\gamma$ dictates
the trade off between the worst case coverage level and the length
of the interval, with a lower level of tolerance $\gamma$ more likely
to provide conservative inference. The feasibility of constructing
the Focused interval relies on the existence of a sufficiently dispersed
distribution of $\Lambda(b')$ at a plausible value $b'$, which may
not exist for extremely low levels of $\gamma$. However, selecting
an extremely low level of $\gamma$ would defeat the purpose of the
Focused interval, which aims to be competitive in terms of \textsl{both}
coverage and length. 

Although Algorithm 1 is novel, the concept of an inference procedure
that depends on a user-specified allowable size distortion has precursors
in the econometrics literature. For example, \citet{Andrews2018}
proposes an inference strategy that controls a worst case coverage
distortion under weak instruments, and where the decision to report
a conventional or weak instrument robust confidence set depends on
the level of under-coverage that an investigator is willing to accept. 

\section{Simulation study}

In this section we illustrate how the finite sample performance of
the Focused estimator and Focused confidence interval depends on the
strength of instruments and the magnitude of direct variant effects
on the outcome. 

First, we consider estimation performance in terms of root-mean squared
error (RMSE). Second, we show how the Focused interval may be able
to achieve a favourable balance of length and coverage, and discuss
where it may lead to improved inference compared with the Core interval,
which is the conventional confidence interval of $\hat{\theta}_{C}$
based on using only the core instruments. Third, we discuss the sensitivity
of the Focused interval to the choice of $\gamma$ which controls
the worst case coverage loss. Finally, we consider the performance
of the Focused estimator when the core instruments $S_{0}$ are in
fact invalid. 

\subsection{Design}

We simulated two-sample summary data on $p=110$ variants according
to Assumptions 1--4. The sample sizes were set at $n=n_{X}=n_{Y}=1000$.
Of the $110$ variants, $10$ were set to be valid instruments, and
they formed the core set $S_{0}$. The remaining $p-\vert S_{0}\vert=100$
variants formed the set of additional instruments $S$. 

We generated estimated associations $\hat{\beta}_{X_{j}}\sim N(\beta_{X_{j}},\sigma_{X_{j}}^{2})$
and $\hat{\beta}_{Y_{j}}\sim N(\beta_{Y_{j}},\sigma_{Y_{j}}^{2})$,
where true genetic variant associations with the risk factor were
set as $\beta_{X_{j}}=\bar{\beta}_{C}\big/\sqrt{\vert S_{0}\vert}$
for $j\in S_{0}$, and $\beta_{X_{j}}=\bar{\beta}_{S}\big/\sqrt{p-\vert S_{0}\vert}$
for $j\in S$, where $\bar{\beta}_{C}$ and $\bar{\beta}_{S}$ were
chosen to maintain a particular level of the \textsl{concentration
parameters} $\lambda_{C}=\sum_{j\in S_{0}}\beta_{X_{j}}^{2}\big/(\vert S_{0}\vert\sigma_{X_{j}}^{2})$
and $\lambda_{S}=\sum_{j\in S}\beta_{X_{j}}^{2}\big/(\vert S\vert\sigma_{X_{j}}^{2})$,
which are measures of the average instrument strength of $S_{0}$
and $S$. The variances $\sigma_{X_{j}}^{2}$ and $\sigma_{Y_{j}}^{2}$
were set equal to $1\big/n$ for all variants. 

The true variant--outcome associations were set to be $\beta_{Y_{j}}=\beta_{X_{j}}\theta_{0}$
for $j\in S_{0}$, and $\beta_{Y_{j}}=\beta_{X_{j}}\theta_{0}+\tau_{j}$
for $j\in S$, where the true causal effect was $\theta_{0}=0.2$,
and the direct effects are fixed effects generated as $\tau_{j}\sim U[0,\bar{\tau}\big/\sqrt{np}]$,
for different values $\bar{\tau}\geq0$. 

For inference, we set the nominal coverage probability at $1-\alpha=0.95$,
and unless otherwise stated, the allowable worst case size distortion
for the Focused interval was set at $\gamma=0.2$. Along with the
Focused, 1-step, 2-step, and Core intervals described above, we also
note the performance of the ``Naive'' confidence interval, which
is the standard confidence interval for the selected estimator that
ignores sampling uncertainty in model selection.

\subsection{Estimation}
\begin{center}
\includegraphics[width=16.5cm]{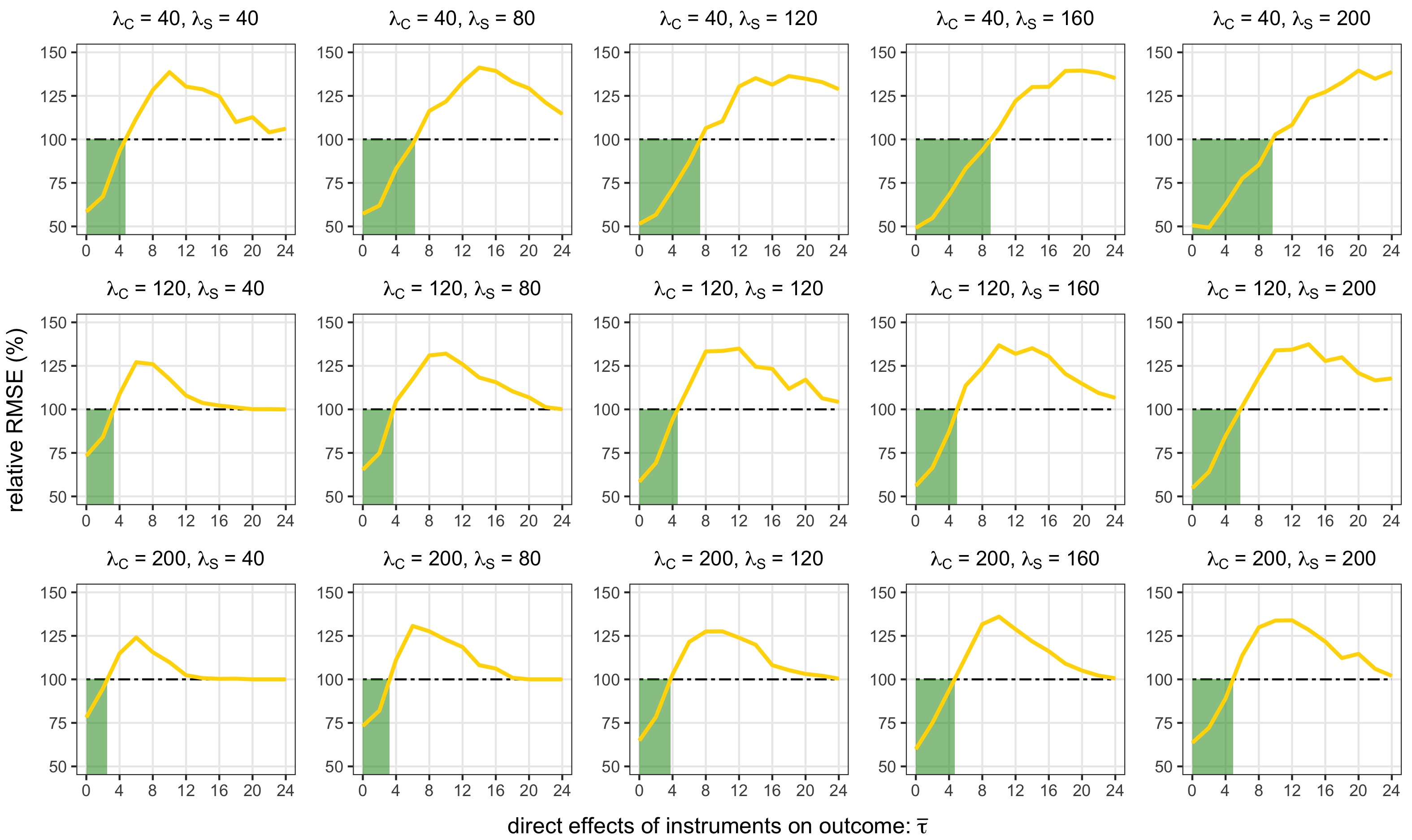}\\
{\small{}Figure 2. }{\footnotesize{}RMSE of Focused estimator relative
to RMSE of Core estimator varying with the average instrument strength
of $S_{0}$ $(\lambda_{C})$ and $S$ $(\lambda_{S})$, and invalidness
of $S$ $(\bar{\tau})$. }{\footnotesize\par}
\par\end{center}

Figure 2 highlights that when the direct variant effects on the outcome
are sufficiently small, the RMSE of the Focused estimator is lower
than the Core estimator. However, this improvement is not uniform
across larger values of $\bar{\tau}$. The performance of the Focused
estimator worsens for more intermediate values of $\bar{\tau}$. Then,
as $\bar{\tau}$ becomes large, the Focused estimator should select
the Core estimator with large probability, so that the RMSEs of the
Core and Focused estimators should be quite similar. 

The extent of the improvement in RMSE that is possible appears to
depend on the strength of instruments. For example, when $\bar{\tau}=2$,
the Focused estimator offered a 32.8\% reduction in RMSE when all
instruments are relatively weak $(\lambda_{C}=\lambda_{S}=40)$, a
30.7\% reduction when $\lambda_{C}=\lambda_{S}=120$, and a 27.8\%
reduction when all instruments are relatively strong $(\lambda_{C}=\lambda_{S}=120)$. 

The relative strengths of the core and additional instrument sets
affect the values of $\bar{\tau}$ over which focused instrument selection
is able to improve estimation. When the additional instruments were
strong and the core instruments were relatively weak $(\lambda_{C}=40$,
$\lambda_{S}=200)$, the Focused estimator had a lower RMSE than the
Core estimator over the range $0\leq\bar{\tau}<10$. In contrast,
when $\lambda_{C}=200$ and $\lambda_{S}=40$, the Focused estimator
had a lower RMSE only over the range $0\leq\bar{\tau}\leq2$. 

In summary, these estimation results intuitively suggest that focused
instrument selection is more likely to improve estimation when the
additional instruments are not too invalid, the core instruments are
quite weak, and the additional instruments are strong. This is practically
relevant for MR analyses of polygenic traits where several genes may
be causally related.

\subsection{Confidence intervals}

For inference, Figure 3 shows that the coverage probability of the
Naive interval dropped to as low as 0.4 when all instruments are equally
strong $(\bar{\tau}=8)$, and lower than 0.2 when the additional instruments
are much stronger than the core instruments $(\lambda_{C}=40,\lambda_{S}=200,\bar{\tau}=10)$.
This underscores the importance for confidence intervals of $\hat{\theta}$
to account for sampling uncertainty in model selection. On the other
hand, the 2-step intervals are conservative; the intervals exceeded
nominal coverage probability, and Figure 4 shows that they were generally
over 30\% longer in length than the Core intervals. 

The 1-step intervals are a useful compromise between the 2-step and
Naive intervals. When the additional instruments were not too invalid
(i.e. for small enough parameter values of $\bar{\tau}$), the 1-step
intervals were shorter in length than the Core intervals, while also
achieving nominal coverage probability. The performance of the 1-step
interval appears to be quite sensitive to the relative strengths of
the core and additional instruments. When the additional instruments
were relatively strong (for example, $\lambda_{C}=40$, $\lambda_{S}\geq160$),
the 1-step intervals were shorter than the Core intervals, but they
under-covered for large enough values of $\bar{\tau}$. Conversely,
when the core instruments are strong enough, the 1-step intervals
showed no real advantages compared with the Core interval. 
\begin{center}
\includegraphics[width=16.5cm]{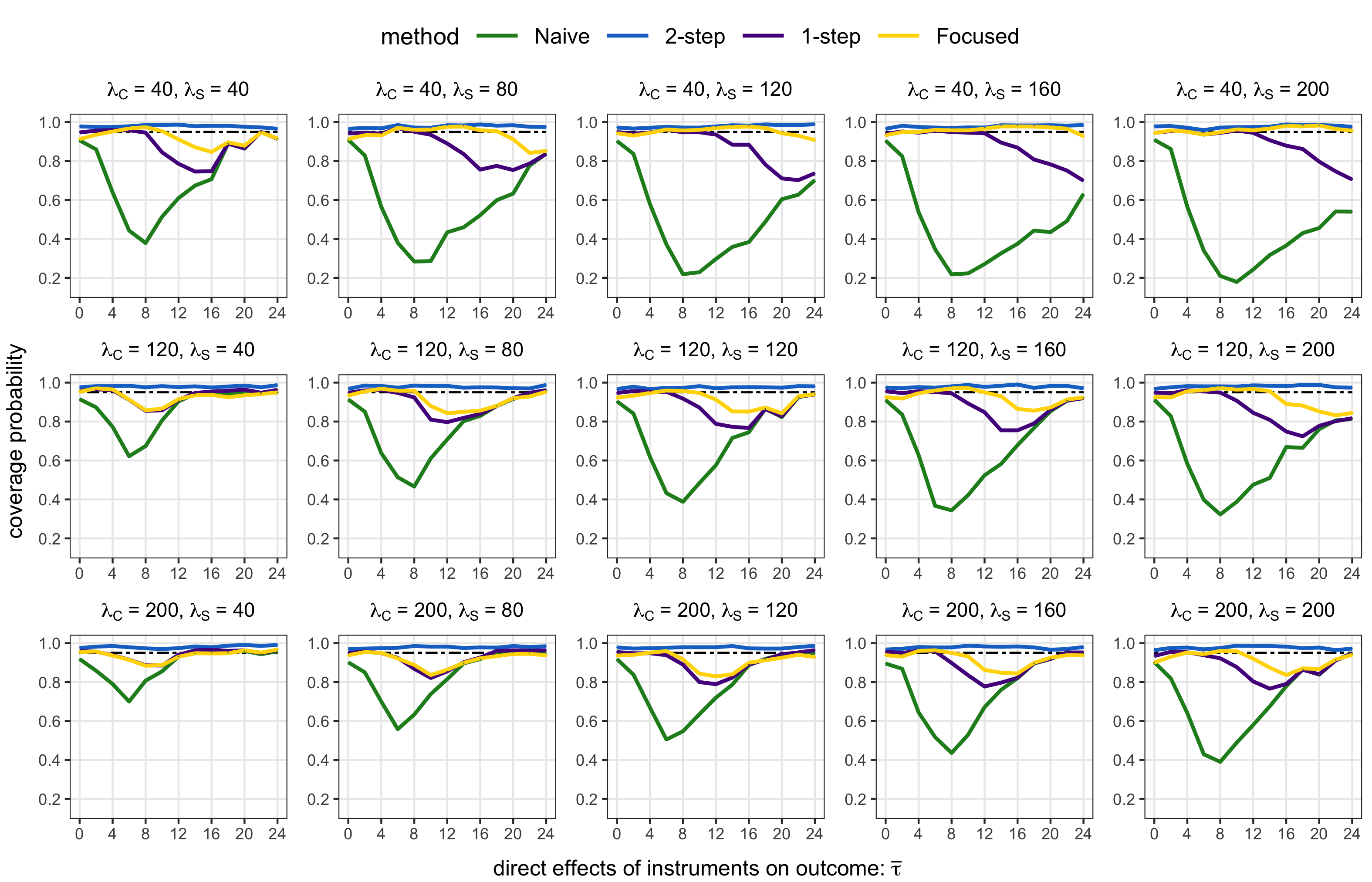}\\
{\small{}Figure 3. }{\footnotesize{}Coverage probabilities of confidence
intervals (nominal coverage is $1-\alpha=0.95$). }{\footnotesize\par}
\par\end{center}

\begin{center}
\includegraphics[width=16.5cm]{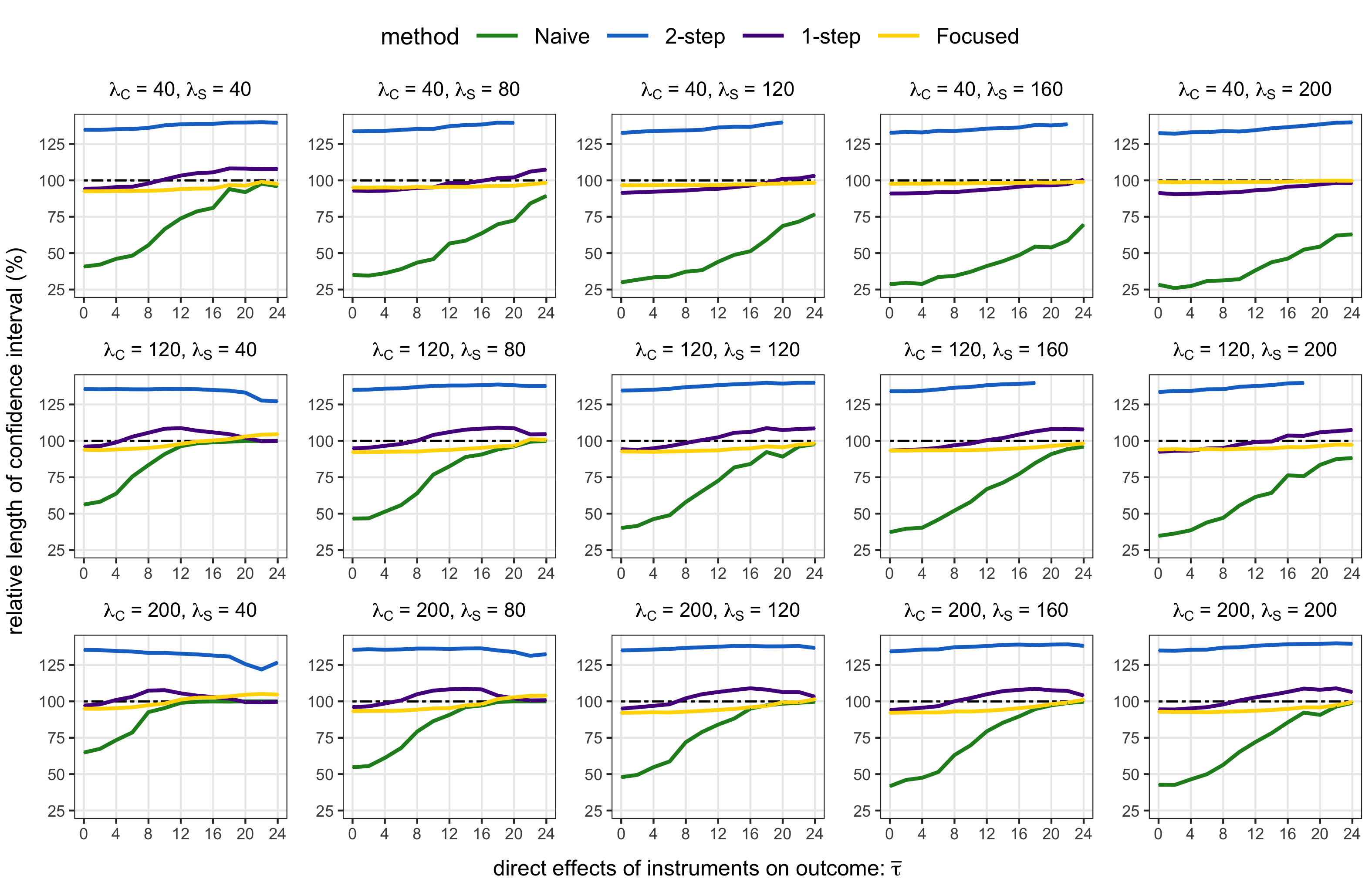}\\
{\small{}Figure 4. }{\footnotesize{}Length of confidence intervals
relative to the Core interval (nominal coverage is $1-\alpha=0.95$). }{\footnotesize\par}
\par\end{center}

For the Focused intervals, we selected the allowable worst case size
distortion as $\gamma=0.2$, so that for a $1-\alpha=0.95$ level
confidence interval, coverage probability should not be lower than
$0.75$. Figure 3 shows that the coverage of the Focused interval
was higher than 0.8 for all parameter values $\bar{\tau}$, whereas
for some values the coverage probability of the 1-step interval dropped
below 0.7. Moreover, the coverage of the Focused interval was generally
more competitive than the 1-step interval, especially when the additional
instruments were at least as strong as the core instruments. 

When the core instruments were not much weaker than the additional
instruments, the Focused intervals were shorter in length than the
1-step interval. Interestingly, the Focused intervals were also up
to 8\% shorter in length than the Core intervals unless the additional
instruments were very invalid $(\bar{\tau}\geq12)$. The Focused interval
can therefore improve the power of an analysis by incorporating information
from additional relevant instruments if they are not too invalid.
At the same time, the Focused interval also retains good size control
and buys insurance against serious under-coverage when additional
instruments are very invalid.

\subsection{Sensitivity to the choice of $\gamma$}

The Focused intervals require investigators to choose an acceptable
level of the worst case size distortion $\gamma$. Figure 5 illustrates
how the performance of the Focused interval varies according to the
choice of $\gamma$ for the case where all instruments are equally
strong $(\lambda_{C}=\lambda_{S}=40)$. The results of other confidence
intervals discussed in this paper are also shown for comparison, but
of course their performance should not vary with $\gamma$. 
\begin{center}
\includegraphics[width=16.5cm]{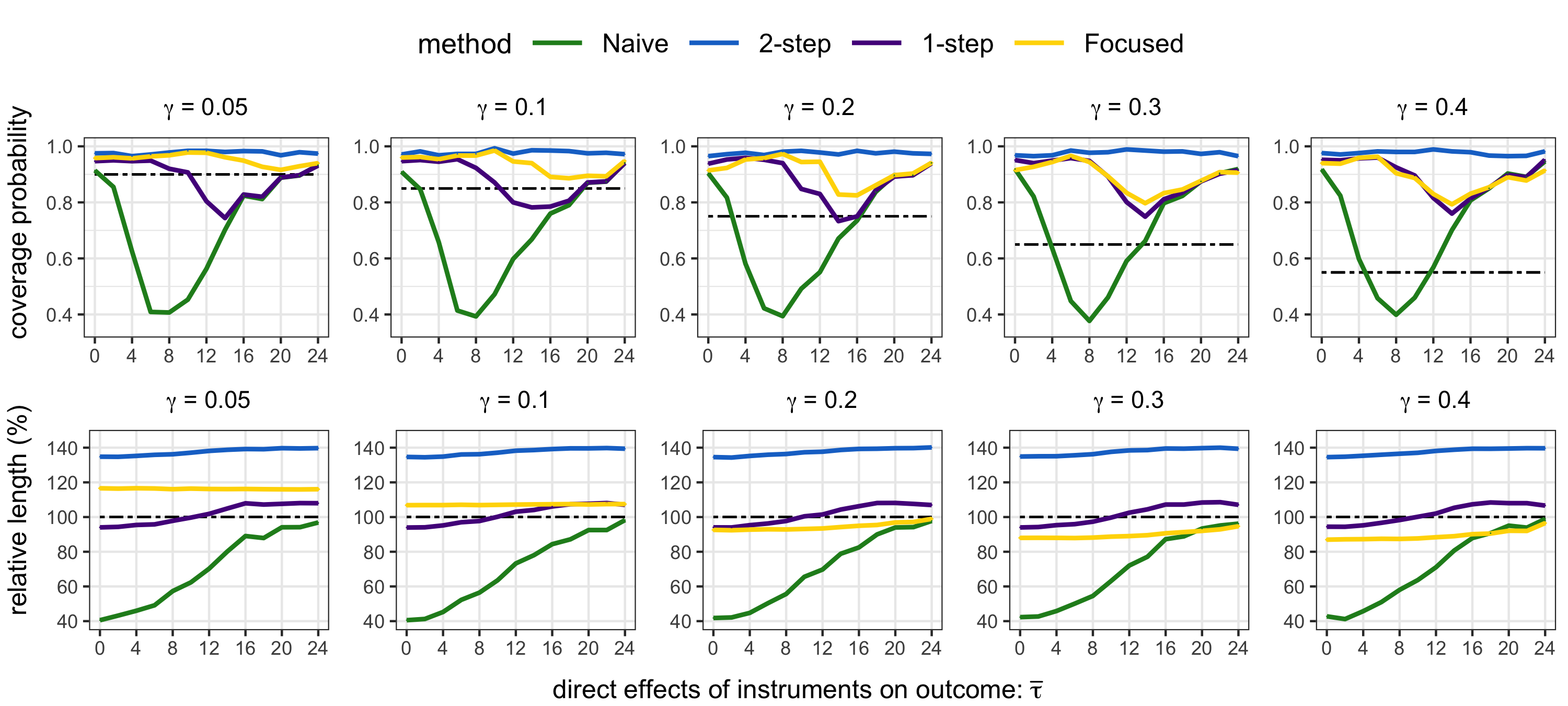}\\
{\small{}Figure 5. }{\footnotesize{}The dashed line in the first row
is the allowable size distortion $1-\alpha-\gamma$ (nominal coverage
is $1-\alpha=0.95$). The second row plots the length of confidence
intervals relative to the Core interval. }{\footnotesize\par}
\par\end{center}

Figure 5 verifies that the Focused interval is able to control the
worst case coverage over all parameter values $\bar{\tau}$. The cost
of allowing only a small size distortion is the longer length of the
interval. The Focused intervals were over 15\% longer in length than
the Core intervals when $\gamma=0.05$, although they were also much
shorter than the 2-step intervals. 

Conversely, for larger values of $\gamma$, the Focused intervals
become shorter, but only to a certain degree: since the Focused intervals
account for uncertainty in model selection, the lengths of the intervals
are not as short as the Naive intervals. Accordingly, the coverage
probability of the Focused intervals will also not change for large
enough values of $\gamma$. 

\subsection{Estimation when $S_{0}$ contains invalid instruments}

Focused instrument selection aims to prioritise the evidence suggested
by a core set of genetic variants $S_{0}$ that are believed to be
valid instruments. In practice, investigators may not always be correct
in their belief, and $S_{0}$ may actually consist of invalid instruments.
We consider this setting in simulation by slightly altering the design
in Section 5.1 so that the true variant--outcome associations are
set to be $\beta_{Y_{j}}=\beta_{X_{j}}\theta_{0}+\tau_{j}$, where
$\tau_{j}\sim U[0,\bar{\tau}_{C}\big/\sqrt{np}]$ for $j\in S_{0}$.
Figure 6 presents the estimation results for the case where all instruments
are equally strong $(\lambda_{C}=\lambda_{S}=40)$. 
\begin{center}
\includegraphics[width=16.5cm]{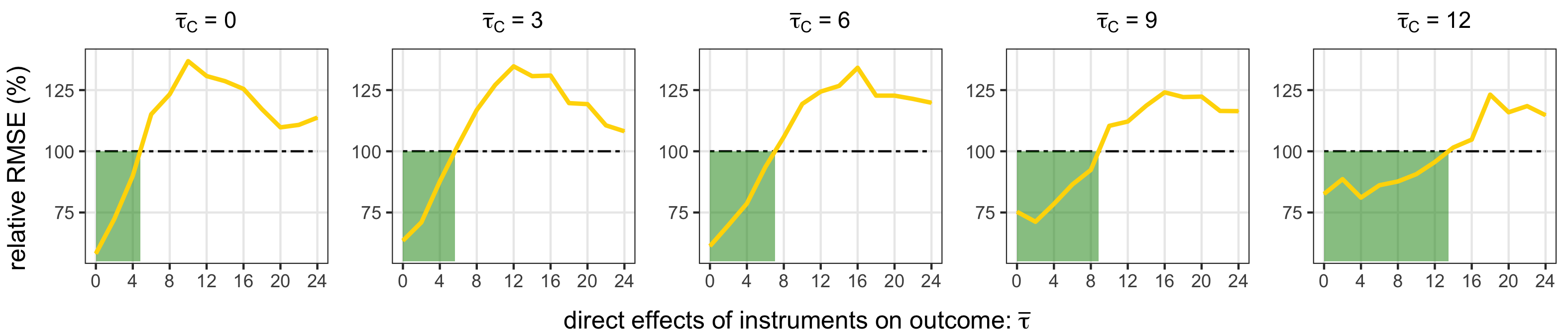}\\
{\small{}Figure 6. }{\footnotesize{}RMSE of Focused estimator relative
to RMSE of Core estimator varying with the invalidness of $S_{0}$
$(\bar{\tau}_{C})$ and invalidness of $S$ $(\bar{\tau})$. }{\footnotesize\par}
\par\end{center}

Our results show that when $S_{0}$ contained only slightly invalid
instruments, the Focused estimator was able to improve on the Core
estimator in terms of RMSE. As the instruments in $S_{0}$ become
more invalid, but not as invalid as $S$, the performance of the Focused
estimator worsens because the estimated bias from including $S$ is
under-estimated. We note that the Focused estimator also performed
relatively well when the instruments $S_{0}$ were at least as invalid
than $S$; for example, the RMSE of the Focused estimator was lower
than that of the Core estimator when $\bar{\tau}_{C}=\bar{\tau}=12$. 

\section{Empirical Examples}

In this section, we demonstrate how focused instrument selection can
be applied in MR studies. First, we consider the problem of instrument
selection in \textsl{drug target MR}, where variation in a gene that
encodes the protein target of a drug is used to proxy drug target
perturbation. Such MR investigations have the potential to provide
genetic evidence on drug efficacy and to study potential side effects
\citep{Gill2021}. 

An important decision in drug target MR is to specify a ``\textsl{cis}
window'' that defines a gene region from which instruments are selected.
In practice, investigators often use \textsl{cis} windows that are
wider than gene regions defined by tools such as GeneCards \citep{Stelzer2016}
in order to boost the power of an analysis through the use of multiple
instruments. This leaves the possibility of a type of publication
bias where researchers may report findings only from a \textsl{cis
w}indow that gives their preferred result. In Section 6.1, we apply
the focused instrument selection method to two lipid drug targets,
where genetic variants that may be included only from widening a \textsl{cis
w}indow are additional instruments.

Second, we investigate the effect of vitamin D supplementation on
a range of outcomes. GWASs have identified strong genetic associations
in biologically plausible genes known to have a functional role in
the transport, synthesis, or metabolism of vitamin D. Some previous
MR studies aiming to study vitamin D effects have used genetic variants
located in neighborhoods of those genes to instrument vitamin D as
they are considered more likely to satisfy the exclusion restriction
than other genome-wide significant variants \citep{Mokry2015,Revez2020}.
However, the role of many other genes which are robustly associated
with vitamin D may not yet be fully understood; for example, \citet{Jiang2021}
selected genetic variants from 69 independent loci to instrument vitamin
D. In Section 6.2, we use genetic variants from biologically plausible
genes as the core set of instruments, and apply focused instrument
selection to select from many additional genetic variants which may
be considered more likely to have a direct effect on the outcome through
their effects on traits other than vitamin D. 

For Focused confidence intervals, we selected $\gamma=0.2$ for the
maximum allowable size distortion. The data used for our analyses
are publicly available through the MR-Base platform \citep{Hemani2020},
and R code to perform our empirical investigation is available on
GitHub at github.com/ash-res/focused-MR/. 

\subsection{CETP and PCSK9 inhibitors }

Cholesteryl ester transfer protein (CETP) inhibitors are a class of
drugs that raise high density lipoprotein cholesterol levels and lower
low density lipoprotein cholesterol (LDL-C) levels. At least three
CETP inhibitors have failed in clinical trials to conclude a protective
effect against coronary heart disease (CHD), but the successful trial
of Anacetrapib showed a modest benefit when used with statins \citep{Bowman2017}.
A recent drug target MR analysis by \citet{Schmidt2021} offers genetic
evidence that CETP inhibition may be an effective approach for preventing
CHD. Here we investigate the robustness of a similar drug target MR
study to the choice of a \textsl{cis} window used to select instruments. 

We study the genetically predicted LDL-C lowering effect of CETP inhibition
on a range of outcomes by using genetic variants located in a neighborhood
of the \textsl{CETP} gene. We may consider instruments drawn from
the ``narrow'' \textsl{cis} window Chr 16: bp: 56,985,862--57,027,757
(which is the region stated in GeneCards $\pm10,000$ bp) to more
accurately represent the genetically predicted effects of CETP inhibition.
At the same time, it is also common for drug target MR studies to
use a ``wider'' \textsl{cis} window for instrument selection. We
may consider potential additional instruments from the wider window
Chr 16: bp: 56,895,862--57,117,757 (which is the region stated in
GeneCards $\pm100,000$ bp). The use of these additional instruments
may be considered more likely to lead to biased estimation compared
with using only those variants from the narrow \textsl{cis} window. 

Therefore the Core estimator used only the 3 uncorrelated genetic
variants located in the narrow window as instruments, while the number
of additional variants that were available varied between 4 and 10
depending on the outcome of interest. The Focused estimator selected
the full set of instruments for 12 out of the 16 outcomes we studied.
The concentration parameter for the core instruments was 91.10, while
the concentration parameter for the additional instruments was between
6.52 and 8.12 depending on how many additional instruments were available
for the application. 
\begin{center}
\includegraphics[width=16.5cm]{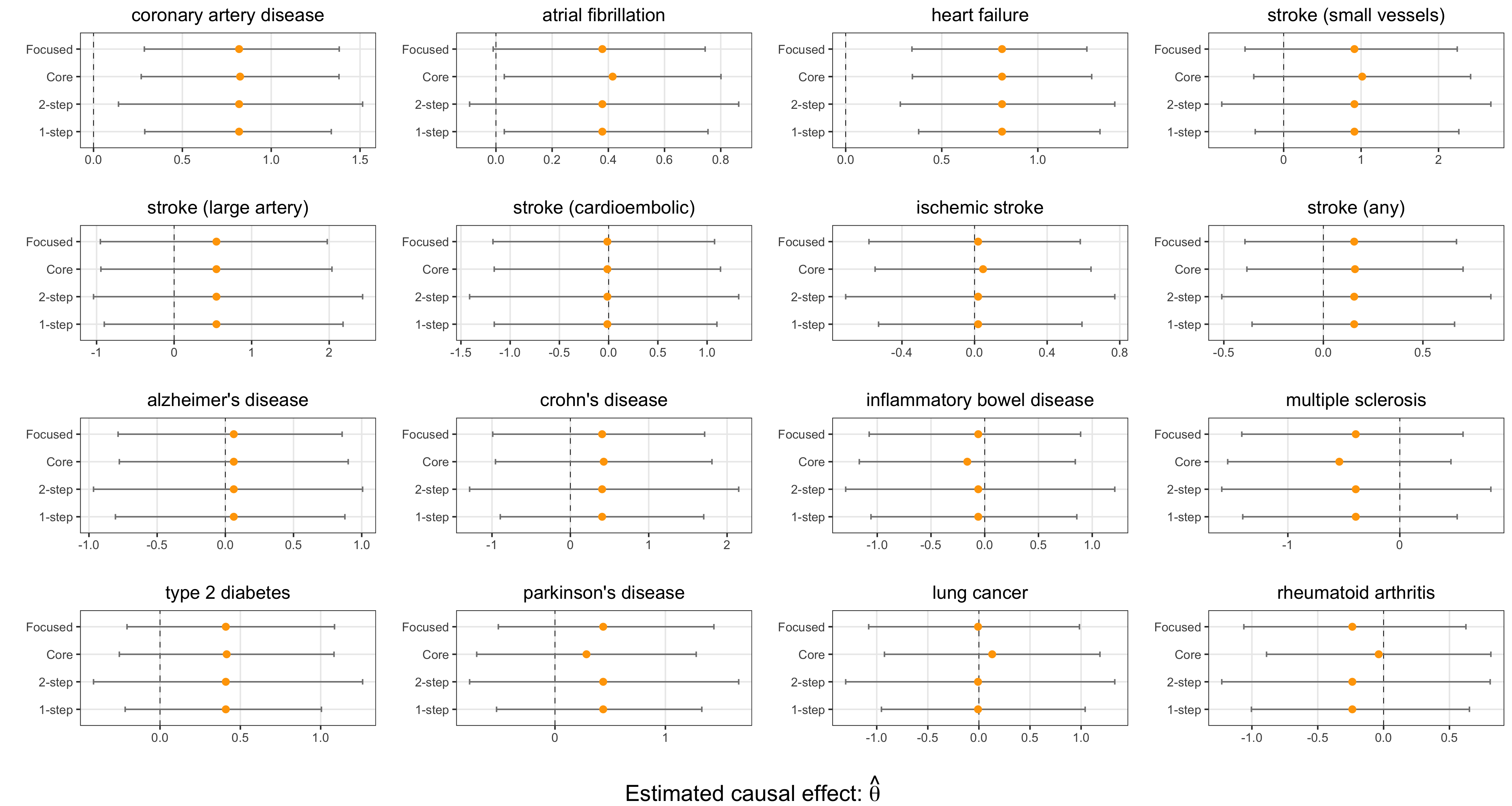}\\
{\small{}Figure 7. }\textsl{\footnotesize{}CETP}{\footnotesize{} gene
analysis. Point estimates and 95\% confidence intervals of the change
in log odds ratio of various outcomes due to a 1 standard deviation
increase in instrumented LDL-C.}{\footnotesize\par}
\par\end{center}

From Figure 7, we find that the genetically predicted LDL-C lowering
effect of CETP inhibition is associated with a lower risk of CHD.
The results also suggest that genetically predicted CETP inhibition
may have a protective effect on other cardiovascular disease outcomes;
specifically atrial fibrillation and heart failure. We do not find
evidence for a protective effect on stroke outcomes, nor do we find
evidence for any adverse effects on various non-cardiovascular related
outcomes. 

Compared with CETP inhibitors, PCSK9 inhibitors (PCSK9i) are a more
established class of drugs that lower LDL-C levels. A drug target
MR analysis can genetically proxy the effect of taking PCSK9i by instrumenting
LDL-C using genetic variants located in a neighborhood of the \textsl{PCSK9}
gene. Similar to the \textsl{CETP} gene analysis above, we consider
variants located in a narrow \textsl{cis} window of \textsl{PCSK9}
(Chr 1: bp: 55,505,221--55,530,525; exactly equal to the window stated
in GeneCards) as core instruments, while additional variants taken
from a wider window $\pm100,000$ bp are additional instruments. 

Using this criteria, there were 4 core instruments when studying the
same outcomes as before, apart from the case of lung cancer where
there were 3. The number of additional instruments available ranged
from 2 to 13. The Focused estimator again selected the full set of
instruments for 12 out of 16 outcomes. The set of core instruments
was very strong compared with the additional instruments; the concentration
parameter for the core instruments was 618.71, compared with a range
of 8.72 to 26.63 for the full set of additional variants available. 
\begin{center}
\includegraphics[width=16.5cm]{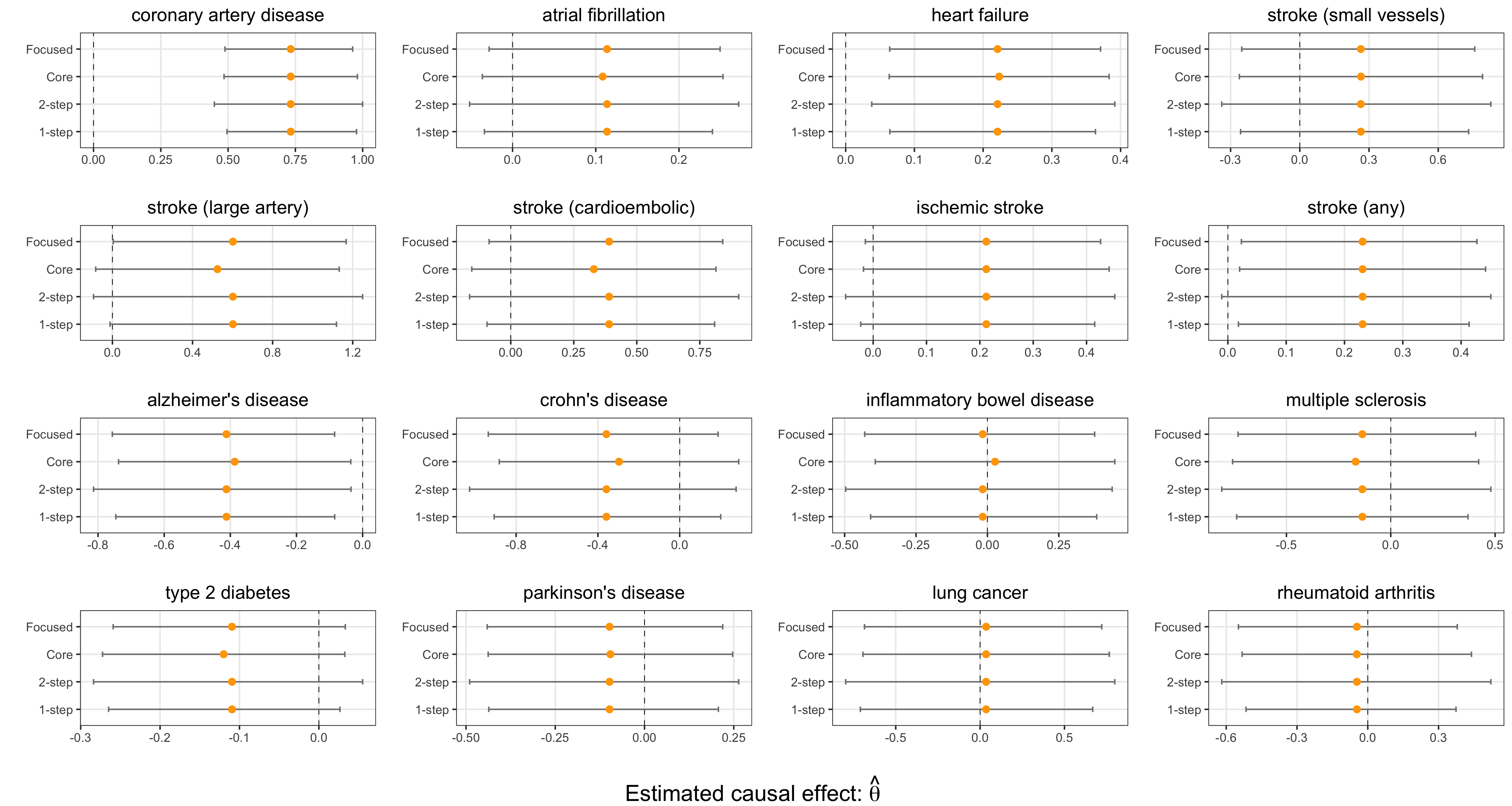}\\
{\small{}Figure 8.}{\footnotesize{} }\textsl{\footnotesize{}PCSK9}{\footnotesize{}
gene analysis. Point estimates and 95\% confidence intervals of the
change in log odds ratio of various outcomes due to a 1 standard deviation
increase in instrumented LDL-C.}{\footnotesize\par}
\par\end{center}

The results in Figure 8 suggest that the genetically predicted LDL-C
lowering effect of PCSK9i is associated with a lower risk of coronary
artery disease, heart failure, and stroke incidence; in addition,
the Focused intervals suggest an association specifically with large
artery stroke incidence. We also find genetic evidence that PCSK9
inhibition may adversely affect the risk of developing Alzheimer's
disease, which supports the findings of \citet{Williams2019} and
\citet{Schmidt2021}. 

\subsection{Vitamin D supplementation}

Finally, we apply focused instrument selection to estimate the genetically
predicted effect of vitamin D supplementation on a range of outcomes.
Previous MR studies instrumenting vitamin D have used variants from
genes implicated in the modulation of 25OHD levels through known mechanisms.
In particular, the \textsl{GC}, \textsl{DHCR7}, \textsl{CYP2R1} and
\textsl{CYP24A1} genes have known functions in vitamin D transport,
synthesis, or metabolism \citep{Berry2012,Mokry2015}. Therefore we
take genetic variants from neighborhoods of these genes ($\pm500,000$
bp on regions stated in GeneCards) as our set of core instruments. 

Moreover, GWASs also provide data on many other robustly associated
genetic variants with vitamin D \citep{Jiang2021}. We use other variants
passing a genome-wide significance threshold (p-value association
with vitamin D less than $5\times10^{-8}$) to form additional instrument
sets. Instead of choosing between the Core and Full estimator, in
our analysis of vitamin D effects we allowed the Focused estimator
to also choose from subsets of the additional instruments. We partitioned
the additional instruments into 3 groups by k-means clustering based
on the ratio estimate of each variant $(\hat{\beta}_{Y_{j}}\big/\hat{\beta}_{X_{j}},j\in S$)
and then considered all possible combinations of these 3 partitions,
thus creating 7 sets of additional instruments. 

In nearly all the outcomes we considered, there were between 10 and
11 genome-wide significant variants from the \textsl{GC}, \textsl{DHCR7},
\textsl{CYP2R1} and \textsl{CYP24A1} gene regions which were used
as core instruments for vitamin D. For the outcomes primary biliary
cirrhosis and asthma there were only 4 variants available to use as
core instruments. Many additional instruments ($\geq50$) were selected
by the Focused estimator in all cases. The core instruments were again
stronger than the additional instruments; for the core instruments
the concentration parameter ranged from 625.52 to 723.21, and for
the additional instruments it ranged from 49.73 to 99.80.

Evidence from observational studies suggests that low serum vitamin
D levels are associated with an increased risk of cardiovascular disease
\citep{Dobnig2008}. These reported associations may be due to unmeasured
confounding, as evidence from a meta-analysis of 21 randomized clinical
trials suggests no causal link \citep{Barbarawi2019}. From Figure
9, we find no genetically predicted effect of vitamin D on cardiovascular
outcomes. 

Our analysis is able to highlight that using the full set of available
instruments can sometimes lead to very different estimates than when
the core instruments are prioritised. In particular, the standard
confidence intervals of the Full estimator suggest a non-null association
of vitamin D on coronary artery disease, heart failure, eczema, primary
biliary cirrhosis, and type 2 diabetes, but these results are not
supported by the Focused intervals.

Our results suggest that higher vitamin D level may have a protective
effect on the incidence of multiple sclerosis, a finding which has
previously been discussed in other MR studies \citep{Mokry2015}.
Through the Core and Focused intervals, we also find that genetically
predicted vitamin D level may be associated with anorexia. 
\begin{center}
\includegraphics[width=16.5cm]{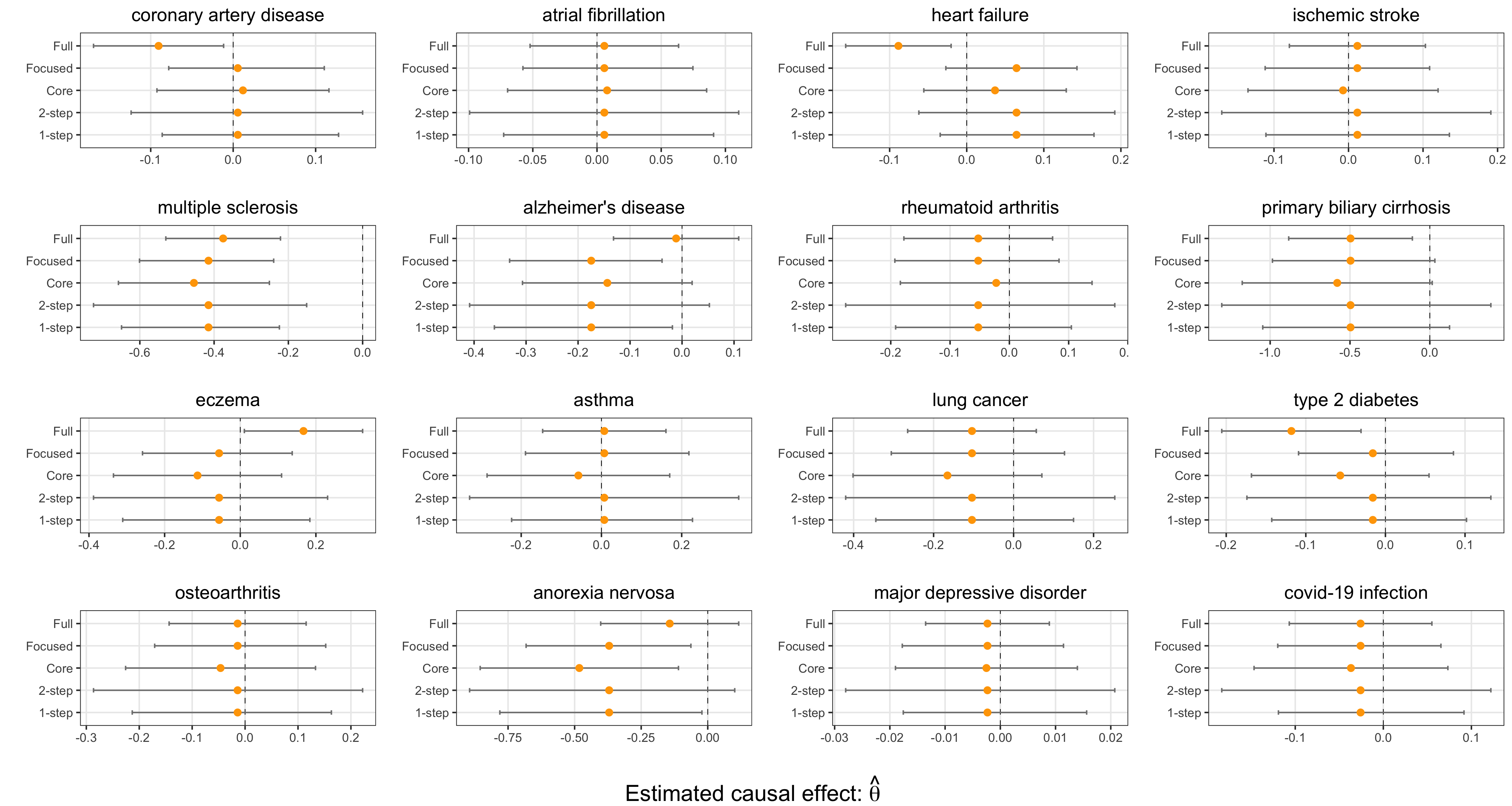}\\
{\small{}Figure 9.}{\footnotesize{} Vitamin D effects. Point estimates
and 95\% confidence intervals of the change in log odds ratio of various
outcomes due to a 1 standard deviation increase in instrumented vitamin
D levels.}{\footnotesize\par}
\par\end{center}

The 1-step and Focused intervals of the Focused estimator suggest
that higher vitamin D level may have a protective effect on the risk
of developing Alzheimer's disease, which supports the findings from
\citet{Jiang2021}, but interestingly this is not supported by the
Core and Full intervals. The Core estimator used 11 instruments, the
Full estimator used 91 additional instruments, and the Focused estimator
selected a subset of 65 of the additional instruments. This illustrates
the ability of the focused instrument selection method to carefully
select additional instruments that can potentially improve the power
of an analysis. 

\section{Conclusion}

Publicly available GWAS summary data have revealed that hundreds of
genetic variants are robustly associated with a wide range of traits
and diseases. However, MR studies in practice do not always make use
of all genetic variants that are strongly associated with risk factors
of interest; in some applications, investigators may have greater
confidence in the instrument validity of only a smaller subset of
many genetic variants. For this setting, we propose a way forward
for improved estimation through \textsl{focused} use of many weak
and potentially invalid instruments. 

Whether focused use of invalid instruments can in turn improve inference
is an open question. While a uniform improvement is not possible,
we propose a new strategy for post-selection inference through Focused
intervals, which are shown to achieve a good balance between precision
and coverage probability while also guarding against a user-specified
worst case size distortion. Our empirical applications highlight the
potential of focused instrument selection to uncover new causal relationships
in MR studies. 

\vskip 3em

\subsection*{Funding acknowledgements}

SB was supported by a Sir Henry Dale Fellowship jointly funded by
the Wellcome Trust and the Royal Society (204623/Z/16/Z). VZ was supported
by the United Kingdom Research and Innovation Medical Research Council
(MR/W029790/1). This research was funded by the United Kingdom Research
and Innovation Medical Research Council (MC-UU-00002/7), and supported
by the National Institute for Health Research Cambridge Biomedical
Research Centre: BRC-1215-20014.

\newpage{}

\appendix

\section{Appendix: Proof of theoretical results}

{\small{}We prove general versions of Theorems 1--4 that allow a
more flexible instrument selection choice. In particular, instead
of simply choosing between the core set of instruments $S_{0}$ and
the full set of instruments, we allow a choice between several additional
instrument sets $S_{1},\ldots,S_{K}$. The results stated in the main
text can be obtained as corollaries for the case $K=1$. Let $S$
denote any generic set of additional instruments from $S_{1},\ldots,S_{K}$,
and let $\hat{\theta}_{S}$ denote the LIML estimator that uses the
set of genetic variants $S_{0}\cup S$ as instruments. }{\small\par}

{\small{}Let CS denote `the Cauchy-Schwarz inequality', CH denote
`Chebyshev's inequality', and T denote `the triangle inequality'.
For all variants $j$, let $e_{Y_{j}}=\hat{\beta}_{Y_{j}}-\beta_{Y_{j}}$,
$e_{X_{j}}=\hat{\beta}_{X_{j}}-\beta_{X_{j}}$, $\hat{g}_{j}(\theta)=\hat{\beta}_{Y_{j}}-\theta\hat{\beta}_{X_{j}}$,
$\Omega_{j}(\theta)=\sigma_{Y_{j}}^{2}+\theta^{2}\sigma_{X_{j}}^{2}$,
and $\Omega_{j}=\Omega_{j}(\theta_{0})$. }\\
{\small{}~}{\small\par}

\textbf{Lemma S.1. (Consistency of $\hat{\theta}_{S}$). }Under Assumptions
1-4, for any additional instrument set $S$, $\hat{\theta}_{S}-\theta_{0}\overset{P}{\to}0$
as $n,p\to\infty$. 

Proof. {\small{}Some simple algebra shows that $\hat{g}_{j}(\theta)=(e_{Y_{j}}-\theta e_{X_{j}})+\beta_{X_{j}}(\theta_{0}-\theta)+\tau_{j}$,
so that for $\hat{Q}(\theta)=-2^{-1}\sum_{j\in S_{0}\cup S}\Omega_{j}(\theta)^{-1}\hat{g}_{j}(\theta)^{2}$,
\begin{eqnarray*}
-2\hat{Q}(\theta) & = & (\theta_{0}-\theta)^{2}\sum_{j\in S_{0}\cup S}\Omega_{j}(\theta)^{-1}\beta_{X_{j}}^{2}+p+\sum_{j\in S_{0}\cup S}\Omega_{j}(\theta)^{-1}[(e_{Y_{j}}-\theta e_{X_{j}})^{2}-\Omega_{j}(\theta)]\\
 &  & +\sum_{j\in S_{0}\cup S}\Omega_{j}(\theta)^{-1}\tau_{j}^{2}+2(\theta_{0}-\theta)\sum_{j\in S_{0}\cup S}\Omega_{j}(\theta)^{-1}\beta_{X_{j}}(e_{Y_{j}}-\theta e_{X_{j}})\\
 &  & +2\sum_{j\in S_{0}\cup S}\Omega_{j}(\theta)^{-1}\tau_{j}(e_{Y_{j}}-\theta e_{X_{j}})+2(\theta_{0}-\theta)\sum_{j\in S_{0}\cup S}\Omega_{j}(\theta)^{-1}\beta_{X_{j}}\tau_{j}\\
 & := & (\theta_{0}-\theta)^{2}\sum_{j\in S_{0}\cup S}\Omega_{j}(\theta)^{-1}\beta_{X_{j}}^{2}+p+R_{1p}+R_{2p}+R_{3p}+R_{4p}+R_{5p}.
\end{eqnarray*}
Note that $E[R_{1p}]=0$ and $Var(R_{1p})=2p$. Hence, $R_{1}=O_{P}(\sqrt{p})$
by CH. Similarly, $R_{3p}=O_{P}(\sqrt{n}\Vert\beta_{X}\Vert_{2}\cdot\vert\theta_{0}-\theta\vert)$,
and $R_{4p}=O_{P}(1)$. By CS, $R_{2p}=O(1)$ and $R_{5p}=O_{P}(\sqrt{n}\Vert\beta_{X}\Vert_{2}\cdot\vert\theta_{0}-\theta\vert)$.
Thus, 
\[
-2\hat{Q}(\theta)=(\theta_{0}-\theta)^{2}\sum_{j\in S_{0}\cup S}\frac{\beta_{X_{j}}^{2}}{\sigma_{Y_{j}}^{2}+\theta^{2}\sigma_{X_{j}}^{2}}+p+O_{P}(\sqrt{p}+\sqrt{n}\Vert\beta_{X}\Vert_{2}\cdot\vert\theta_{0}-\theta\vert).
\]
The rest of the proof is then identical to the Proof of Theorem 3.1
of \citet[p.41]{Zhao2018}. ~\hfill{}$\square$}\\
{\small{}~}{\small\par}

For Lemma S.2-S.4, let $\psi_{j}(\theta)=\Omega_{j}(\theta)^{-2}(\hat{\beta}_{Y_{j}}-\theta\hat{\beta}_{X_{j}})(\hat{\beta}_{X_{j}}\sigma_{Y_{j}}^{2}+\theta\hat{\beta}_{Y_{j}}\sigma_{X_{j}}^{2})$,
for each variant $j$. 

~

\textbf{Lemma S.2.} Under Assumptions 1-4, $\sum_{j\in S_{0}\cup S}\nabla_{\theta}\psi_{j}(\theta_{0})\big/(\eta_{C}+\eta_{S})\overset{P}{\to}-1$
as $n,p\to\infty$. 

Proof. {\small{}The first-order condition is given by $\sum_{j\in S_{0}\cup S}\psi_{j}(\hat{\theta}_{S})=0$.
Also, 
\begin{eqnarray*}
\nabla_{\theta}\psi_{j}(\theta) & = & -\hat{\beta}_{X_{j}}(\hat{\beta}_{X_{j}}\sigma_{Y_{j}}^{2}+\theta\hat{\beta}_{Y_{j}}\sigma_{X_{j}}^{2})\Omega_{j}(\theta)^{-2}+\hat{\beta}_{Y_{j}}\sigma_{X_{j}}^{2}(\hat{\beta}_{Y_{j}}-\theta\hat{\beta}_{X_{j}})\Omega_{j}(\theta)^{-2}\\
 &  & -4\theta\sigma_{X_{j}}^{2}(\hat{\beta}_{Y_{j}}-\theta\hat{\beta}_{X_{j}})(\hat{\beta}_{X_{j}}\sigma_{Y_{j}}^{2}+\theta\hat{\beta}_{Y_{j}}\sigma_{X_{j}}^{2})\Omega_{j}(\theta)^{-3}
\end{eqnarray*}
where
\[
E[\nabla_{\theta}\psi_{j}(\theta_{0})]=-\Omega_{j}^{-1}\beta_{X_{j}}^{2}+\Omega_{j}^{-2}\sigma_{X_{j}}^{2}\tau_{j}^{2}-4\theta_{0}\Omega_{j}^{-2}\sigma_{X_{j}}^{2}\beta_{X_{j}}\tau_{j}-4\theta_{0}^{2}\Omega_{j}^{-3}\sigma_{X_{j}}^{4}\tau_{j}^{2}.
\]
Let $\phi_{j}(\theta)=\nabla_{\theta}\psi_{j}(\theta)-E[\nabla_{\theta}\psi_{j}(\theta)]$.
Then, 
\begin{eqnarray*}
\phi_{j}(\theta_{0}) & = & -\Omega_{j}^{-1}\beta_{X_{j}}e_{X_{j}}-\Omega_{j}^{-2}\sigma_{Y_{j}}^{2}\beta_{X_{j}}e_{X_{j}}-\theta_{0}\Omega_{j}^{-2}\sigma_{X_{j}}^{2}\beta_{X_{j}}e_{Y_{j}}-\Omega_{j}^{-2}\sigma_{Y_{j}}^{2}(e_{X_{j}}^{2}-\sigma_{X_{j}}^{2})\\
 &  & -\theta_{0}\Omega_{j}^{-2}\sigma_{X_{j}}^{2}\tau_{j}e_{X_{j}}-\theta_{0}\Omega_{j}^{-2}\sigma_{X_{j}}^{2}e_{Y_{j}}e_{X_{j}}+\theta_{0}\Omega_{j}^{-2}\sigma_{X_{j}}^{2}\beta_{X_{j}}e_{Y_{j}}-\theta_{0}^{2}\Omega_{j}^{-2}\sigma_{X_{j}}^{2}\beta_{X_{j}}e_{X_{j}}\\
 &  & +\Omega_{j}^{-2}\sigma_{X_{j}}\tau_{j}e_{Y_{j}}-\theta_{0}\Omega_{j}^{-2}\sigma_{X_{j}}^{2}\tau_{j}e_{X_{j}}+\Omega_{j}^{-2}\sigma_{X_{j}}^{2}(e_{Y_{j}}^{2}-\sigma_{Y_{j}}^{2})-\theta_{0}\Omega_{j}^{-2}\sigma_{X_{j}}^{2}e_{Y_{j}}e_{X_{j}}\\
 &  & +\Omega_{j}^{-2}\sigma_{X_{j}}^{2}\tau_{j}e_{Y_{j}}-4\theta_{0}\Omega_{j}^{-2}\sigma_{X_{j}}^{2}\beta_{X_{j}}e_{Y_{j}}-4\theta_{0}\Omega_{j}^{-3}\sigma_{X_{j}}^{2}\sigma_{Y_{j}}^{2}e_{Y_{j}}e_{X_{j}}-4\theta_{0}^{2}\Omega_{j}^{-3}\sigma_{X_{j}}^{4}\tau_{j}e_{Y_{j}}\\
 &  & -4\theta_{0}^{2}\Omega_{j}^{-3}\sigma_{X_{j}}^{4}(e_{Y_{j}}^{2}-\sigma_{Y_{j}}^{2})+4\theta_{0}^{2}\Omega_{j}^{-2}\sigma_{X_{j}}^{2}\beta_{X_{j}}e_{X_{j}}+4\theta_{0}^{2}\Omega_{j}^{-3}\sigma_{Y_{j}}^{2}\sigma_{X_{j}}^{2}(e_{X_{j}}^{2}-\sigma_{X_{j}}^{2})\\
 &  & +4\theta_{0}^{3}\Omega_{j}^{-3}\sigma_{X_{j}}^{4}\tau_{j}e_{X_{j}}+4\theta_{0}^{3}\Omega_{j}^{-3}\sigma_{X_{j}}^{4}e_{Y_{j}}e_{X_{j}}-4\theta_{0}\Omega_{j}^{-3}\sigma_{Y_{j}}^{2}\sigma_{X_{j}}^{2}\tau_{j}e_{X_{j}}-4\theta_{0}^{2}\Omega_{j}^{-3}\sigma_{X_{j}}^{4}\tau_{j}e_{Y_{j}}\\
 &  & \equiv\sum_{l=1}^{23}\phi_{j}^{(l)}(\theta_{0}).
\end{eqnarray*}
First, $Var(\phi_{j}^{(1)}(\theta_{0}))=\Omega_{j}^{-2}\beta_{X_{j}}^{2}\sigma_{X_{j}}^{2}$,
so that $\sum_{j\in S_{0}\cup S}Var(\phi_{j}^{(1)}(\theta_{0}))=\Theta(n\Vert\beta_{X}\Vert_{2}^{2})$.
Similarly, for each $l\in[23]$, we have that $\sum_{j\in S_{0}\cup S}Var(\phi_{j}^{(l)}(\theta_{0}))=O(n\Vert\beta_{X}\Vert_{2}^{2})+O(p)$.
Therefore, by CH, 
\begin{eqnarray*}
P\Big(\Big\vert\frac{1}{\eta_{C}+\eta_{S}}\sum_{j\in S_{0}\cup S}\phi_{j}(\theta_{0})\Big\vert>\kappa\Big) & \leq & \frac{1}{\kappa^{2}(\eta_{C}+\eta_{S})^{2}}\Big(\sum_{j\in S_{0}\cup S}Var(\phi_{j}(\theta_{0}))+2\sum_{j\in S_{0}\cup S}\sum_{k\neq j}Cov(\phi_{j}(\theta_{0}),\phi_{k}(\theta_{0}))\Big)\\
 & \leq & O(1\big/n\Vert\beta_{X}\Vert_{2}^{2})+O(p\big/n^{2}\Vert\beta_{X}\Vert_{2}^{4})\\
 & = & o(1),
\end{eqnarray*}
where the second inequality uses $Cov(\phi_{j}(\theta_{0}),\phi_{k}(\theta_{0}))\leq Var(\phi_{j}(\theta_{0}))^{\frac{1}{2}}Var(\phi_{k}(\theta_{0}))^{\frac{1}{2}}$
and $\eta_{C}+\eta_{S}=\Theta(n\Vert\beta_{X}\Vert_{2}^{2})$, and
the equality follows by Assumption 3. Therefore,
\[
\frac{1}{\eta_{C}+\eta_{S}}\sum_{j\in S_{0}\cup S}\nabla_{\theta}\psi_{j}(\theta_{0})=\frac{1}{\eta_{C}+\eta_{S}}\sum_{j\in S_{0}\cup S}E[\nabla_{\theta}\psi_{j}(\theta_{0})]+o_{P}(1).
\]
}{\small\par}

{\small{}and 
\begin{eqnarray*}
\frac{1}{\eta_{C}+\eta_{S}}\sum_{j\in S_{0}\cup S}E[\nabla_{\theta}\psi_{j}(\theta_{0})] & = & -1+\frac{1}{\eta_{C}+\eta_{S}}\Big[\sum_{j\in S_{0}\cup S}\Omega_{j}^{-2}\sigma_{X_{j}}^{2}\tau_{j}^{2}-4\theta_{0}\sum_{j\in S_{0}\cup S}\Omega_{j}^{-2}\sigma_{X_{j}}^{2}\beta_{X_{j}}\tau_{j}\\
 &  & -4\theta_{0}^{2}\sum_{j\in S_{0}\cup S}\Omega_{j}^{-3}\sigma_{X_{j}}^{4}\tau_{j}^{2}\Big]\\
 & = & -1+\frac{1}{\eta_{C}+\eta_{S}}[O(n\Vert\tau\Vert_{2}^{2})+O(n\Vert\beta_{X}\Vert_{2}\Vert\tau\Vert_{2})]\\
 & = & -1+O(1\big/n\Vert\beta_{X}\Vert_{2}^{2})+O(1\big/\sqrt{n}\Vert\beta_{X}\Vert_{2})\\
 & = & -1+o_{P}(1).
\end{eqnarray*}
Hence, by T, $\sum_{j\in S_{0}\cup S}\nabla_{\theta}\psi_{j}(\theta_{0})\big/(\eta_{C}+\eta_{S})=-1+o_{P}(1)$.
\hfill{}$\square$}\\
{\small{}~}{\small\par}

\textbf{Lemma S.3. }Under Assumptions 1-4, 
\[
\sum_{j\in S_{0}\cup S}\psi_{j}(\theta_{0})\big/\sqrt{\eta_{C}+\eta_{S}+\varsigma_{C}+\varsigma_{S}}\overset{D}{\to}N(b_{S}/\sqrt{\eta_{C}+\eta_{S}+\varsigma_{C}+\varsigma_{S}},1)
\]
 as $n,p\to\infty$.

Proof. {\small{}We can decompose $\psi_{j}(\theta_{0})$ into a fixed
bias term $b_{j}$, and stochastic terms $J_{1j}$ and $J_{2j}$,
\[
\psi_{j}(\theta_{0})=b_{j}+J_{1j}+J_{2j},
\]
where $b_{j}=\Omega_{j}^{-1}\beta_{X_{j}}\tau_{j}+\theta_{0}\Omega_{j}^{-2}\sigma_{X_{j}}^{2}\tau_{j}^{2}$,
$J_{1j}=\Omega_{j}^{-1}\beta_{X_{j}}(e_{Y_{j}}-\theta_{0}e_{X_{j}})+\Omega_{j}^{-2}(e_{Y_{j}}-\theta_{0}e_{X_{j}})(\sigma_{Y_{j}}^{2}e_{X_{j}}+\theta_{0}\sigma_{X_{j}}^{2}e_{Y_{j}})$,
$J_{2j}=\Omega_{j}^{-2}(\sigma_{Y_{j}}^{2}e_{X_{j}}+2\theta_{0}\sigma_{X_{j}}^{2}e_{Y_{j}}-\theta_{0}^{2}\sigma_{X_{j}}^{2}e_{X_{j}})\tau_{j}$. }{\small\par}

{\small{}By CS, 
\[
\sum_{j\in S_{0}\cup S}\Omega_{j}^{-2}\sigma_{X_{j}}^{2}\tau_{j}^{2}=\Theta(n\Vert\tau\Vert_{2}^{2})=O(1).
\]
Also, 
\begin{eqnarray*}
Var\Big(\sum_{j\in S_{0}\cup S}J_{2j}\Big) & = & \sum_{j\in S_{0}\cup S}\Omega_{j}^{-4}\tau_{j}^{2}\sigma_{X_{j}}^{2}\big[(\sigma_{Y_{j}}^{2}-\theta_{0}^{2}\sigma_{X_{j}}^{2})^{2}+4\theta_{0}^{2}\sigma_{X_{j}}^{2}\sigma_{Y_{j}}^{2}\big]\\
 & = & \sum_{j\in S_{0}\cup S}\Omega_{j}^{-2}\tau_{j}^{2}\sigma_{X_{j}}^{2}\\
 & = & O(1),
\end{eqnarray*}
where the last equality follows by $n\Vert\tau\Vert_{2}^{2}=O(1)$
by Assumption 4. Therefore, by CH and $E[J_{2j}]=0$ for all $j$,
$\sum_{j\in S_{0}\cup S}J_{2j}=O_{P}(1)$. }{\small\par}

{\small{}Then, since $\eta_{C}+\eta_{S}+\varsigma_{C}+\varsigma_{S}=\Theta(n\Vert\beta_{X}\Vert_{2}^{2}+p)$,
\[
\frac{1}{\sqrt{\eta_{C}+\eta_{S}+\varsigma_{C}+\varsigma_{S}}}\sum_{j\in S_{0}\cup S}\big[\psi_{j}(\theta_{0})-\Omega_{j}^{-1}\beta_{X_{j}}\tau_{j}\big]=\Big(\frac{1}{\sqrt{\eta_{C}+\eta_{S}+\varsigma_{C}+\varsigma_{S}}}\sum_{j\in S_{0}\cup S}J_{1j}\Big)+o_{P}(1).
\]
By identical arguments used in \citet{Zhao2018}, $(\eta_{C}+\eta_{S}+\varsigma_{C}+\varsigma_{S})^{-\frac{1}{2}}\sum_{j\in S_{0}\cup S}J_{1j}\overset{D}{\to}N(0,1)$.
The result then follows by Slutsky's lemma. \hfill{}$\square$}{\small\par}

{\small{}~}{\small\par}

\textbf{Lemma S.4. }Under Assumptions 1-4, for any $\bar{\theta}\overset{P}{\to}\theta_{0}$,
$\sum_{j\in S_{0}\cup S}\nabla_{\theta\theta}\psi_{j}(\bar{\theta})\big/(\eta_{C}+\eta_{S})=O_{P}(1)$
as $n,p\to\infty$. 

Proof. {\small{}Note that $\nabla_{\theta}\Omega_{j}(\theta)=2\theta\sigma_{X_{j}}^{2}$.
For any $\theta$, we can write 
\begin{eqnarray*}
\nabla_{\theta\theta}\psi_{j}(\theta) & = & -2\hat{\beta}_{X_{j}}\hat{\beta}_{Y_{j}}\sigma_{X_{j}}^{2}\Omega_{j}(\theta)^{-2}+2\hat{\beta}_{X_{j}}(\hat{\beta}_{X_{j}}\sigma_{Y_{j}}^{2}+\theta\hat{\beta}_{Y_{j}}\sigma_{X_{j}}^{2})(\nabla_{\theta}\Omega_{j}(\theta))\Omega_{j}(\theta)^{-3}\\
 &  & -2\hat{\beta}_{Y_{j}}\sigma_{X_{j}}^{2}(\hat{\beta}_{Y_{j}}-\theta\hat{\beta}_{X_{j}})(\nabla_{\theta}\Omega_{j}(\theta))\Omega_{j}(\theta)^{-3}\\
 &  & -4\sigma_{X_{j}}^{2}(\hat{\beta}_{Y_{j}}-\theta\hat{\beta}_{X_{j}})(\hat{\beta}_{X_{j}}\sigma_{Y_{j}}^{2}+\theta\hat{\beta}_{Y_{j}}\sigma_{X_{j}}^{2})\Omega_{j}(\theta)^{-3}\\
 &  & +4\theta\sigma_{X_{j}}^{2}\hat{\beta}_{X_{j}}(\hat{\beta}_{X_{j}}\sigma_{Y_{j}}^{2}+\theta\hat{\beta}_{Y_{j}}\sigma_{X_{j}}^{2})\Omega_{j}(\theta)^{-3}-4\theta\sigma_{X_{j}}^{4}\hat{\beta}_{Y_{j}}(\hat{\beta}_{Y_{j}}-\theta\hat{\beta}_{X_{j}})\Omega_{j}(\theta)^{-3}\\
 &  & +12\theta\sigma_{X_{j}}^{2}(\hat{\beta}_{Y_{j}}-\theta\hat{\beta}_{X_{j}})(\hat{\beta}_{X_{j}}\sigma_{Y_{j}}^{2}+\theta\hat{\beta}_{Y_{j}}\sigma_{X_{j}}^{2})(\nabla_{\theta}\Omega_{j}(\theta))\Omega_{j}(\theta)^{-4}\\
 & \equiv & \sum_{l=1}^{8}H_{lj}(\theta).
\end{eqnarray*}
We can expand $H_{1j}(\theta)$ so that 
\begin{eqnarray*}
\sum_{j\in S_{0}\cup S}H_{1j}(\theta) & = & -\sum_{j\in S_{0}\cup S}(\beta_{X_{j}}+e_{X_{j}})(\theta_{0}\beta_{X_{j}}+e_{Y_{j}}+\tau_{j})\sigma_{X_{j}}^{2}\Omega_{j}(\theta)^{-2}\\
 & = & -\theta_{0}\sum_{j\in S_{0}\cup S}\sigma_{X_{j}}^{2}\Omega_{j}(\theta)^{-2}\beta_{X_{j}}^{2}-\sum_{j\in S_{0}\cup S}\sigma_{X_{j}}^{2}\Omega_{j}(\theta)^{-2}\beta_{X_{j}}e_{Y_{j}}-\sum_{j\in S_{0}\cup S}\sigma_{X_{j}}^{2}\Omega_{j}(\theta)^{-2}\beta_{X_{j}}\tau_{j}\\
 &  & -\theta_{0}\sum_{j\in S_{0}\cup S}\sigma_{X_{j}}^{2}\Omega_{j}(\theta)^{-2}\beta_{X_{j}}e_{X_{j}}-\sum_{j\in S_{0}\cup S}\sigma_{X_{j}}^{2}\Omega_{j}(\theta)^{-2}e_{X_{j}}e_{Y_{j}}-\sum_{j\in S_{0}\cup S}\sigma_{X_{j}}^{2}\Omega_{j}(\theta)^{-2}e_{X_{j}}\tau_{j}.
\end{eqnarray*}
For the first term on the right hand side, note that $\sum_{j\in S_{0}\cup S}\sigma_{X_{j}}^{2}\Omega_{j}(\theta)^{-2}\beta_{X_{j}}^{2}=\Theta(n\Vert\beta_{X}\Vert_{2}^{2})$
since $\Omega_{j}(\theta)=\Theta(n^{-1})$. For the second term, }{\small\par}

{\small{}
\[
Var\big(\sum_{j\in S_{0}\cup S}\sigma_{X_{j}}^{2}\Omega_{j}(\theta)^{-2}\beta_{X_{j}}e_{Y_{j}}\big)=\sum_{j\in S_{0}\cup S}\sigma_{X_{j}}^{4}\sigma_{Y_{j}}^{2}\Omega_{j}(\theta)^{-4}\beta_{X_{j}}^{2}=\Theta(n\Vert\beta_{X}\Vert_{2}^{2}),
\]
so that by CH, $\sum_{j\in S_{0}\cup S}\sigma_{X_{j}}^{2}\Omega_{j}(\theta)^{-2}\beta_{X_{j}}e_{Y_{j}}=O(\sqrt{n}\Vert\beta_{X}\Vert_{2})$.}{\small\par}

{\small{}Similarly,
\[
\Big\vert\sum_{j\in S_{0}\cup S}\sigma_{X_{j}}^{2}\Omega_{j}(\theta)^{-2}\beta_{X_{j}}\tau_{j}\Big\vert\leq\Theta(n\Vert\beta_{X}\Vert_{2}\Vert\tau\Vert_{2})=O(\sqrt{n}\Vert\beta_{X}\Vert_{2}).
\]
}{\small\par}

{\small{}For the fifth term on the right hand side, 
\[
Var\big(\sum_{j\in S_{0}\cup S}\sigma_{X_{j}}^{2}\Omega_{j}(\theta)^{-2}e_{X_{j}}e_{Y_{j}}\big)=\Theta\Big(n^{2}\sum_{j\in S_{0}\cup S}\sigma_{X_{j}}^{2}\sigma_{Y_{j}}^{2}\Big)=\Theta(p),
\]
so that $\sum_{j\in S_{0}\cup S}\sigma_{X_{j}}^{2}\Omega_{j}(\theta)^{-2}e_{X_{j}}e_{Y_{j}}=O(\sqrt{p})$
by CH. }{\small\par}

{\small{}Using similar arguments for the remaining terms of $\sum_{j\in S_{0}\cup S}H_{1j}(\theta)$,
and for $\sum_{j\in S_{0}\cup S}H_{lj}(\theta)$, $l=2,\ldots,8$,
we have that
\[
\sum_{l=1}^{8}\sum_{j\in S_{0}\cup S}H_{lj}(\theta)=O_{P}(n\Vert\beta_{X}\Vert_{2}^{2}+p)
\]
as $n,p\to\infty$, which leads to the required result since $\eta_{C}+\eta_{S}=\Theta(n\Vert\beta_{X}\Vert_{2}^{2})$
and $p\big/(n\Vert\beta_{X}\Vert_{2}^{2})=O(1)$ by Assumption 3.
\hfill{}$\square$}{\small\par}

{\small{}~}{\small\par}

\textbf{Proof of Theorem 1 (Asymptotic distribution of $\hat{\theta}_{S}$).}

{\small{}The proof strategy follows }\textcolor{blue}{\small{}\citet{Zhao2018}'s}{\small{}
Proof of Theorem 3.2, but with Lemmas S.2-S.4 involving additional
steps required to control the non-negligible asymptotic bias term
from our model. }{\small\par}

{\small{}Let $\psi_{j}(\theta)=(\hat{\beta}_{Y_{j}}-\theta\hat{\beta}_{X_{j}})(\hat{\beta}_{X_{j}}\sigma_{Y_{j}}^{2}+\theta\hat{\beta}_{Y_{j}}\sigma_{X_{j}}^{2})(\sigma_{Y_{j}}^{2}+\theta^{2}\sigma_{X_{j}}^{2})^{-2}$.
Given consistency of $\hat{\theta}_{S}$, a second-order Taylor expansion
of the first-order condition $\sum_{j\in S_{0}\cup S}\psi_{j}(\hat{\theta}_{S})=0$
around $\hat{\theta}_{S}=\theta_{0}$ implies that there exists $\bar{\theta}$
on the line segment joining $\hat{\theta}_{S}$ and $\theta_{0}$
such that 
\begin{eqnarray*}
\frac{\eta_{C}+\eta_{S}}{\sqrt{\eta_{C}+\eta_{S}+\varsigma_{C}+\varsigma_{S}}}(\hat{\theta}_{S}-\theta_{0}) & = & -\Big(\frac{1}{\eta_{C}+\eta_{S}}\sum_{j\in S_{0}\cup S}\nabla_{\theta}\psi_{j}(\theta_{0})+o_{P}\big(\frac{1}{\eta_{C}+\eta_{S}}\sum_{j\in S_{0}\cup S}\nabla_{\theta\theta}\psi_{j}(\bar{\theta})\big)\Big)^{-1}\\
 &  & \times\frac{1}{\sqrt{\eta_{C}+\eta_{S}+\varsigma_{C}+\varsigma_{S}}}\sum_{j\in S_{0}\cup S}\psi_{j}(\theta_{0}).
\end{eqnarray*}
The result then follows by Slutsky's lemma, and Lemmas S.2-S.4, which
show, }as $n,p\to\infty${\small{}:}{\small\par}

\textbf{\small{}(i)}{\small{} $\sum_{j\in S_{0}\cup S}\nabla_{\theta}\psi_{j}(\theta_{0})\big/(\eta_{C}+\eta_{S})\overset{P}{\to}-1;$}\\
\textbf{\small{}(ii)}{\small{} $\sum_{j\in S_{0}\cup S}\psi_{j}(\theta_{0})\big/\sqrt{\eta_{C}+\eta_{S}+\varsigma_{C}+\varsigma_{S}}\overset{D}{\to}N(b_{S}/\sqrt{\eta_{C}+\eta_{S}+\varsigma_{C}+\varsigma_{S}},1)$;}\\
\textbf{\small{}(iii)}{\small{} for any $\bar{\theta}\overset{P}{\to}\theta_{0}$,
$\sum_{j\in S_{0}\cup S}\nabla_{\theta\theta}\psi_{j}(\bar{\theta})\big/(\eta_{C}+\eta_{S})=O_{P}(1)$.
\hfill{}$\square$}{\small\par}

{\small{}~}{\small\par}

\textbf{Proof of Theorem 2 (Asymptotic distribution of the bias $\hat{b}_{S}$).}

{\small{}We show that $\hat{V}_{B}^{-\frac{1}{2}}(\hat{b}_{S}-b_{S})\overset{D}{\to}N(0,1)$
as $n,p\to\infty$, where $\hat{V}_{B}=\hat{\eta}_{S}+\hat{\varsigma}_{S}+\hat{\xi}_{S}+(\hat{\eta}_{S}^{2}\big/\hat{\eta}_{C}^{2})(\hat{\eta}_{C}+\hat{\varsigma}_{C})$.
The result of Theorem 2 then follows by $(\hat{\eta}_{C}+\hat{\eta}_{S})/(\eta_{C}+\eta_{S})\overset{P}{\to}1$
shown below in Proof of Lemma S.5, and Slutsky's lemma. }{\small\par}

{\small{}Let $\hat{B}_{j}(\theta)=\hat{\Omega}_{j}^{-1}\hat{\beta}_{X_{j}}(\hat{\beta}_{Y_{j}}-\theta\hat{\beta}_{X_{j}})+\theta\hat{\Omega}_{j}^{-1}\sigma_{X_{j}}^{2}$.
Note that $\hat{\Omega}_{j}-\Omega_{j}=(\hat{\theta}-\theta_{0})(\hat{\theta}+\theta_{0})\sigma_{X_{j}}^{2}=\Theta(n^{-1}\vert\hat{\theta}-\theta_{0}\vert)$
implies that $\hat{\Omega}_{j}^{-1}-\Omega_{j}^{-1}=\hat{\Omega}_{j}^{-1}(\Omega_{j}-\hat{\Omega}_{j})\Omega_{j}^{-1}=\Theta(n\vert\hat{\theta}-\theta_{0}\vert)$. }{\small\par}

{\small{}We can write 
\begin{eqnarray*}
\sum_{j\in S}\big[\hat{B}_{j}(\theta_{0})-\Omega_{j}^{-1}\beta_{X_{j}}\tau_{j}\big] & = & \sum_{j\in S}\bigg[\Omega_{j}^{-1}\beta_{X_{j}}(e_{Y_{j}}-\theta_{0}e_{X_{j}})+\Omega_{j}^{-1}e_{X_{j}}e_{Y_{j}}-\theta_{0}\Omega_{j}^{-1}(e_{X_{j}}^{2}-\sigma_{X_{j}}^{2})\\
 &  & +(\hat{\Omega}_{j}^{-1}-\Omega_{j}^{-1})\beta_{X_{j}}(e_{Y_{j}}-\theta_{0}e_{X_{j}})+(\hat{\Omega}_{j}^{-1}-\Omega_{j}^{-1})e_{X_{j}}e_{Y_{j}}\\
 &  & -\theta_{0}(\hat{\Omega}_{j}^{-1}-\Omega_{j}^{-1})(e_{X_{j}}^{2}-\sigma_{X_{j}}^{2})+(\hat{\Omega}_{j}^{-1}-\Omega_{j}^{-1})\beta_{X_{j}}\tau_{j}\\
 &  & +\Omega_{j}^{-1}\tau_{j}e_{X_{j}}+(\hat{\Omega}_{j}^{-1}-\Omega_{j}^{-1})\tau_{j}e_{X_{j}}\bigg].
\end{eqnarray*}
Using $\hat{\Omega}_{j}^{-1}-\Omega_{j}^{-1}=\Theta(n\vert\hat{\theta}-\theta_{0}\vert)$,
note that $\big\vert\sum_{j\in S}(\hat{\Omega}_{j}^{-1}-\Omega_{j}^{-1})\beta_{X_{j}}(e_{Y_{j}}-\theta_{0}e_{X_{j}})\big\vert=O_{P}(\sqrt{n}\vert\hat{\theta}-\theta_{0}\vert\Vert\beta_{X}\Vert_{2})=o_{P}(\sqrt{n}\Vert\beta_{X}\Vert_{2})$
by CS, CH, and consistency of $\hat{\theta}_{C}$ for $\theta_{0}$.
Similarly, the last five terms on the right hand side are $o_{P}(\sqrt{n}\Vert\beta_{X}\Vert_{2}+\sqrt{p})$.
Therefore, for $\bar{B}_{j}(\theta_{0})=\Omega_{j}^{-1}\beta_{X_{j}}(e_{Y_{j}}-\theta_{0}e_{X_{j}})+\Omega_{j}^{-1}e_{X_{j}}e_{Y_{j}}-\theta_{0}\Omega_{j}^{-1}(e_{X_{j}}^{2}-\sigma_{X_{j}}^{2})$,
\[
\sum_{j\in S}\big[\hat{B}_{j}(\theta_{0})-\Omega_{j}^{-1}\beta_{X_{j}}\tau_{j}\big]=\sum_{j\in S}\bar{B}_{j}(\theta_{0})+o_{P}(\sqrt{n}\Vert\beta_{X}\Vert_{2}+\sqrt{p}).
\]
By CH, $\vert e_{X_{j}}\vert=O_{P}(n^{-\frac{1}{2}})$ and $\vert e_{Y_{j}}\vert=O_{P}(n^{-\frac{1}{2}})$,
so that $E[\vert\bar{B}_{j}(\theta_{0})\vert^{3}]=O(n^{\frac{3}{2}}\vert\beta_{X_{j}}\vert^{3})+O(n\vert\beta_{X_{j}}\vert^{2})+O(\sqrt{n}\vert\beta_{X_{j}}\vert)+O(1)$,
and 
\begin{eqnarray*}
\sum_{j\in S}E\big[\vert\bar{B}_{j}(\theta_{0})\vert^{3}\big] & = & O(n^{\frac{3}{2}}\Vert\beta_{X}\Vert_{3}^{3})+O(n\Vert\beta_{X}\Vert_{2}^{2})+O(\sqrt{n}\Vert\beta_{X}\Vert_{1})+O(p)\\
 & = & O(n^{\frac{3}{2}}\Vert\beta_{X}\Vert_{3}^{3})+O(n\Vert\beta_{X}\Vert_{2}^{2})+O(p),
\end{eqnarray*}
where the last equality follows by $\sqrt{n}\Vert\beta_{X}\Vert_{1}\leq\sqrt{p}\sqrt{n}\Vert\beta_{X}\Vert_{2}\leq(n\Vert\beta_{X}\Vert_{2}^{2}+p)\big/2$. }{\small\par}

{\small{}Let $\xi_{S}=2\theta_{0}^{2}\sum_{j\in S}\Omega_{j}^{-2}\sigma_{X_{j}}^{4}$.
The variance of $\sum_{j\in S}\bar{B}_{j}(\theta_{0})$ is given by
\[
Var\Big(\sum_{j\in S}\bar{B}_{j}(\theta_{0})\Big)=\eta_{S}+\varsigma_{S}+\xi_{S},
\]
where $\eta_{S}+\varsigma_{S}=\Theta(n\Vert\beta_{X}\Vert_{2}^{2}+p)$
and $\xi_{S}=\Theta(p)$. Therefore, the following Lyapanov condition
holds, 
\begin{eqnarray*}
\frac{1}{(\eta_{S}+\varsigma_{S}+\xi_{S})^{\frac{3}{2}}}\sum_{j\in S}E\big[\vert\bar{B}_{j}(\theta_{0})\vert^{3}\big] & = & O\Big(\frac{\Vert\beta_{X}\Vert_{3}}{\Vert\beta_{X}\Vert_{2}}\Big)+O\Big(\frac{1}{\sqrt{n}\Vert\beta_{X}\Vert_{2}}\Big)+O\Big(\frac{1}{\sqrt{p}}\Big)\\
 & = & o(1),
\end{eqnarray*}
by $\Vert\beta_{X}\Vert_{3}\big/\Vert\beta_{X}\Vert_{2}\to0$ in Assumption
3. Thus, by Lyapanov's CLT, 
\[
\frac{1}{\sqrt{\eta_{S}+\varsigma_{S}+\xi_{S}}}\sum_{j\in S}\bar{B}_{j}(\theta_{0})\overset{D}{\to}N(0,1).
\]
}{\small\par}

{\small{}Note that $\hat{B}_{j}(\hat{\theta}_{C})-\hat{B}_{j}(\theta_{0})=-(\hat{\theta}_{C}-\theta_{0})\hat{\Omega}_{j}^{-1}(\hat{\beta}_{X_{j}}^{2}-\sigma_{X_{j}}^{2})$,
and 
\begin{eqnarray*}
\sum_{j\in S}\hat{\Omega}_{j}^{-1}(\hat{\beta}_{X_{j}}^{2}-\sigma_{X_{j}}^{2}) & = & \sum_{j\in S}\bigg[\Omega_{j}^{-1}\beta_{X_{j}}^{2}+(\hat{\Omega}_{j}^{-1}-\Omega_{j}^{-1})\beta_{X_{j}}^{2}+2\Omega_{j}^{-1}\beta_{X_{j}}e_{X_{j}}+2(\hat{\Omega}_{j}^{-1}-\Omega_{j}^{-1})\beta_{X_{j}}e_{X_{j}}\\
 &  & +\Omega_{j}^{-1}(e_{X_{j}}^{2}-\sigma_{X_{j}}^{2})+(\hat{\Omega}_{j}^{-1}-\Omega_{j}^{-1})(e_{X_{j}}^{2}-\sigma_{X_{j}}^{2})\bigg].
\end{eqnarray*}
By similar arguments used above, and since $p\big/n\Vert\beta_{X}\Vert_{2}^{2}=O(1)$
by Assumption 3, the last five terms on the right hand side are $o_{P}(n\Vert\beta_{X}\Vert_{2}^{2})$.}{\small\par}

{\small{}Therefore, since $\hat{\theta}_{C}-\theta_{0}=O_{P}(1\big/\sqrt{n}\Vert\beta_{X,0}\Vert_{2})+O(\sqrt{p}\big/n\Vert\beta_{X,0}\Vert_{2}^{2})$,
by the above results,
\[
\sum_{j\in S}\big[\hat{B}_{j}(\hat{\theta}_{C})-\Omega_{j}^{-1}\beta_{X_{j}}\tau_{j}\big]=\sum_{j\in S}\bar{B}_{j}(\theta_{0})-\eta_{S}(\hat{\theta}_{C}-\theta_{0})+o_{P}(\sqrt{n}\Vert\beta_{X}\Vert_{2}+\sqrt{p}).
\]
Let $V_{B}=\eta_{S}+\varsigma_{S}+\xi_{S}+(\eta_{S}^{2}\big/\eta_{C}^{2})(\eta_{C}+\varsigma_{C})$.
Note that $V_{B}=\Theta(n\Vert\beta_{X}\Vert_{2}^{2}+p)$ so that
\[
V_{B}^{-\frac{1}{2}}\sum_{j\in S}\big[\hat{B}_{j}(\hat{\theta}_{C})-\Omega_{j}^{-1}\beta_{X_{j}}\tau_{j}\big]=V_{B}^{-\frac{1}{2}}\Big(\sum_{j\in S}\bar{B}_{j}(\theta_{0})-\eta_{S}(\hat{\theta}_{C}-\theta_{0})\Big)+o_{P}(1).
\]
It can be shown that the Core estimator has the first order expansion
\[
\hat{\theta}_{C}-\theta_{0}=\frac{1}{\eta_{C}}\sum_{j\in S_{0}}J_{1j}+o_{P}\Big(\frac{\sqrt{\eta_{C}+\varsigma_{C}}}{\eta_{C}}\Big),
\]
so that 
\[
V_{B}^{-\frac{1}{2}}\sum_{j\in S}\big[\hat{B}_{j}(\hat{\theta}_{C})-\Omega_{j}^{-1}\beta_{X_{j}}\tau_{j}\big]=V_{B}^{-\frac{1}{2}}\Big(\sum_{j\in S}\bar{B}_{j}(\theta_{0})-\frac{\eta_{S}}{\eta_{C}}\sum_{j\in S_{0}}J_{1j}\Big)+o_{P}(1),
\]
as $o_{P}(V_{B}^{-\frac{1}{2}}(\eta_{S}\big/\eta_{C}^{2})\sqrt{\eta_{C}+\zeta_{C}})=o_{P}(1)$.
From the above arguments, and noting that $\sum_{j\in S}\bar{B}_{j}(\theta_{0})$
and $\sum_{j\in S_{0}}J_{1j}$ are mutually independent, the result
follows by Slutsky's lemma.\hfill{}$\square$}{\small\par}

{\small{}~}{\small\par}

\textbf{Lemma S.5 (Consistent variance estimation).}

{\small{}Let $\hat{\Omega}_{j}=\sigma_{Y_{j}}^{2}+\hat{\theta}_{C}^{2}\sigma_{X_{j}}^{2}$,
$\hat{V}_{B}=\hat{\eta}_{S}+\hat{\varsigma}_{S}+\hat{\xi}_{S}+(\hat{\eta}_{S}^{2}\big/\hat{\eta}_{C}^{2})(\hat{\eta}_{C}+\hat{\varsigma}_{C})$,
where $\hat{\xi}_{S}=2\hat{\theta}_{C}^{2}\sum_{j\in S}\hat{\Omega}_{j}^{-2}\sigma_{X_{j}}^{4}$. }{\small\par}

{\small{}We show that, under Assumptions 1-4, }as $n,p\to\infty$, 

\textbf{\small{}(i)}{\small{} $(\eta_{C}+\eta_{S})^{2}(\hat{\eta}_{C}+\hat{\eta}_{S}+\hat{\varsigma}_{C}+\hat{\varsigma}_{S})\big/(\hat{\eta}_{C}+\hat{\eta}_{S})^{2}(\eta_{C}+\eta_{S}+\varsigma_{C}+\varsigma_{S})\overset{P}{\to}1$;}\\
\textbf{\small{}(ii)}{\small{} $\hat{V}_{B}\big/V_{B}\overset{P}{\to}1$. }{\small\par}

\textsl{Proof. }

\textbf{\small{}Part (i)}{\small\par}

{\small{}First, note that $\hat{\Omega}_{j}-\Omega_{j}=2\sigma_{X_{j}}^{2}(\hat{\theta}_{S}-\theta_{0})=\Theta(\vert\hat{\theta}_{S}-\theta_{0}\vert\big/n)$,
so that $\hat{\Omega}_{j}^{-1}-\Omega_{j}^{-1}=\hat{\Omega}_{j}^{-1}(\Omega_{j}-\hat{\Omega}_{j})\Omega_{j}^{-1}=\Theta(n\vert\hat{\theta}_{S}-\theta_{0}\vert)$.}{\small\par}

{\small{}Then, }{\small\par}

{\small{}
\begin{eqnarray*}
\hat{\eta}_{C}+\hat{\eta}_{S}-\eta_{C}-\eta_{S} & = & 2\sum_{j\in S_{0}\cup S}\Omega_{j}^{-1}\beta_{X_{j}}e_{X_{j}}+\sum_{j\in S_{0}\cup S}\Omega_{j}^{-1}(e_{X_{j}}^{2}-\sigma_{X_{j}}^{2})+\sum_{j\in S_{0}\cup S}(\hat{\Omega}_{j}^{-1}-\Omega_{j}^{-1})\beta_{X_{j}}^{2}\\
 &  & +2\sum_{j\in S_{0}\cup S}(\hat{\Omega}_{j}^{-1}-\Omega_{j}^{-1})\beta_{X_{j}}e_{X_{j}}+\sum_{j\in S_{0}\cup S}(\hat{\Omega}_{j}^{-1}-\Omega_{j}^{-1})(e_{X_{j}}^{2}-\sigma_{X_{j}}^{2})\\
 & = & O_{P}(\sqrt{n}\Vert\beta_{X}\Vert_{2})+O_{P}(\sqrt{p})+o_{P}(n\Vert\beta_{X}\Vert_{2}^{2})\\
 & = & o_{P}(n\Vert\beta_{X}\Vert_{2}^{2})+O_{P}(\sqrt{p}),
\end{eqnarray*}
by similar arguments used in the proof of Theorem 2. By T, since $\eta_{C}+\eta_{S}=\Theta(n\Vert\beta_{X}\Vert_{2}^{2})$
and $p\big/n^{2}\Vert\beta_{X}\Vert_{2}^{4}\to0$, we have $(\hat{\eta}_{C}+\hat{\eta}_{S})^{-1}=O_{P}(1\big/n\Vert\beta_{X}\Vert_{2}^{2})$. }{\small\par}

{\small{}Therefore, 
\begin{eqnarray*}
\frac{(\eta_{C}+\eta_{S})^{2}}{(\hat{\eta}_{C}+\hat{\eta}_{S})^{2}}-1 & = & 2(\eta_{C}+\eta_{S}+\hat{\eta}_{C}+\hat{\eta}_{S})(\eta_{C}+\eta_{S}-\hat{\eta}_{C}-\hat{\eta}_{S})\big/(\hat{\eta}_{C}+\hat{\eta}_{S})^{2}\\
 & = & O_{P}(n\Vert\beta_{X}\Vert_{2}^{2})[o_{P}(n\Vert\beta_{X}\Vert_{2}^{2})+O_{P}(\sqrt{p})]O_{P}(1\big/n^{2}\Vert\beta_{X}\Vert_{2}^{4})\\
 & = & o_{P}(1)+O_{P}(\sqrt{p}\big/n\Vert\beta_{X}\Vert_{2}^{2})\\
 & = & o_{P}(1),
\end{eqnarray*}
where the last line follows from $p\big/n^{2}\Vert\beta_{X}\Vert_{2}^{4}\to0$
as $n,p\to\infty$, which is implied by Assumption 3.}{\small\par}

{\small{}Similarly, noting that $\hat{\Omega}_{j}^{-2}-\Omega_{j}^{-2}=\hat{\Omega}_{j}^{-2}\Omega_{j}^{-2}(\Omega_{j}+\hat{\Omega}_{j})(\Omega_{j}-\hat{\Omega}_{j})=\Theta(n^{2}\vert\hat{\theta}_{S}-\theta_{0}\vert)$,
we have 
\begin{eqnarray*}
\hat{\varsigma}_{C}+\hat{\varsigma}_{S}-\varsigma_{C}-\varsigma_{S} & = & \sum_{j\in S_{0}\cup S}(\hat{\Omega}_{j}^{-2}-\Omega_{j}^{-2})\sigma_{X_{j}}^{2}\sigma_{Y_{j}}^{2}\\
 & = & \Theta(p\vert\hat{\theta}_{S}-\theta_{0}\vert)\\
 & = & o_{P}(p).
\end{eqnarray*}
}{\small\par}

{\small{}Using the above results,
\begin{eqnarray*}
\frac{(\eta_{C}+\eta_{S})^{2}}{(\hat{\eta}_{C}+\hat{\eta}_{S})^{2}}\cdot\frac{\hat{\eta}_{C}+\hat{\eta}_{S}+\hat{\varsigma}_{C}+\hat{\varsigma}_{S}}{\eta_{C}+\eta_{S}+\varsigma_{C}+\varsigma_{S}} & = & (1+o_{P}(1))\Big(1+\frac{\hat{\eta}_{C}+\hat{\eta}_{S}-\eta_{C}-\eta_{S}}{\eta_{C}+\eta_{S}+\varsigma_{C}+\varsigma_{S}}+\frac{\hat{\varsigma}_{C}+\hat{\varsigma}_{S}-\varsigma_{C}-\varsigma_{S}}{\eta_{C}+\eta_{S}+\varsigma_{C}+\varsigma_{S}}\Big)\\
 & = & (1+o_{P}(1))\Big(1+\frac{o_{P}(n\Vert\beta_{X}\Vert_{2}^{2})+O_{P}(\sqrt{p})+o_{P}(p)}{\Theta(n\Vert\beta_{X}\Vert_{2}^{2})+\Theta(p)}\Big)\\
 & = & (1+o_{P}(1))^{2}\\
 & = & 1+o_{P}(1).
\end{eqnarray*}
}{\small\par}

\textbf{\small{}Part (ii)}{\small\par}

{\small{}First, note that 
\begin{eqnarray*}
\hat{\xi}_{S}-\xi_{S} & = & 2(\hat{\theta}_{S}+\theta_{0})(\hat{\theta}_{S}-\theta_{0})\sum_{j\in S}\Omega_{j}^{-2}\sigma_{X_{j}}^{4}+2(\hat{\theta}_{S}+\theta_{0})(\hat{\theta}_{S}-\theta_{0})\sum_{j\in S}(\hat{\Omega}_{j}^{-2}-\Omega_{j}^{-2})\sigma_{X_{j}}^{4}\\
 &  & +2\theta_{0}^{2}\sum_{j\in S}(\hat{\Omega}_{j}^{-2}-\Omega_{j}^{-2})\sigma_{X_{j}}^{4}\\
 & = & \Theta(p\vert\hat{\theta}_{S}-\theta_{0}\vert)+\Theta(p\vert\hat{\theta}_{S}-\theta_{0}\vert^{2})\\
 & = & o_{P}(p).
\end{eqnarray*}
Therefore, 
\begin{eqnarray*}
\hat{V}_{B}-V_{B} & = & (\hat{\eta}_{S}-\eta_{S})+(\hat{\varsigma}_{S}-\varsigma_{S})+(\hat{\xi}_{S}-\xi_{S})+(\hat{\eta}_{S}-\eta_{S})(\hat{\eta}_{S}+\eta_{S})\frac{(\hat{\eta}_{C}+\hat{\zeta}_{V})}{\hat{\eta}_{C}^{2}}\\
 &  & +\eta_{S}^{2}\Big(\frac{\hat{\eta}_{C}+\hat{\zeta}_{C}}{\hat{\eta}_{C}^{2}}-\frac{\eta_{C}+\zeta_{C}}{\eta_{C}^{2}}\Big)\\
 & = & o_{P}(n\Vert\beta_{X}\Vert_{2}^{2})+O_{P}(\sqrt{p})+o_{P}(p)+\big(o_{P}(n\Vert\beta_{X}\Vert_{2}^{2})+O_{P}(\sqrt{p})\big)O_{P}(n\Vert\beta_{X}\Vert_{2}^{2})\big[O_{P}(1\big/n\Vert\beta_{X}\Vert_{2}^{2})\\
 &  & +o_{P}(1)\big]+O(n^{2}\Vert\beta_{X}\Vert_{2}^{4})\big(o_{P}(1/n\Vert\beta_{X}\Vert_{2}^{2})+o_{P}(p/n^{2}\Vert\beta_{X}\Vert_{2}^{4})\big)\\
 & = & o_{P}(n\Vert\beta_{X}\Vert_{2}^{2})+o_{P}(p)+o_{P}(p^{2}\big/n\Vert\beta_{X}\Vert_{2}^{2})
\end{eqnarray*}
where the second equality follows by (i) and T, since $\text{\ensuremath{\hat{\eta}_{C}^{-2}}}(\hat{\eta}_{C}+\hat{\varsigma}_{C})=O_{P}(\eta_{C}^{-2}(\eta_{C}+\varsigma_{C}))=O_{P}(1\big/n\Vert\beta_{X}\Vert_{2}^{2})+O_{P}(p\big/n^{2}\Vert\beta_{X}\Vert_{2}^{4})$,
and $\text{\ensuremath{\hat{\eta}_{C}^{-2}}}(\hat{\eta}_{C}+\hat{\varsigma}_{C})-\eta_{C}^{-2}(\eta_{C}+\varsigma_{C})=o_{P}(\eta_{C}^{-2}(\eta_{C}+\varsigma_{C}))=o_{P}(1/n\Vert\beta_{X}\Vert_{2}^{2})+o_{P}(p/n^{2}\Vert\beta_{X}\Vert_{2}^{4})$. }{\small\par}

{\small{}Therefore, since $V_{B}=\Theta(n\Vert\beta_{X}\Vert_{2}^{2})+\Theta(p)$,
\begin{align*}
\frac{\hat{V}_{B}-V_{B}}{V_{B}} & =o_{P}(1)+o_{P}\Big(\frac{p}{n\Vert\beta_{X}\Vert_{2}^{2}}\Big)\\
 & =o_{P}(1),
\end{align*}
as $p\big/n\Vert\beta_{X}\Vert_{2}^{2}=O(1)$. Finally, we note $\hat{V}_{B}\big/V_{B}=1+(\hat{V}_{B}-V_{B})\big/V_{B}=1+o_{P}(1)$.
\hfill{}$\square$}\\
{\small{}~}{\small\par}

\textbf{Proof of Theorem 3 -- Part I (Convergence in distribution
of effect and bias estimates).} 

{\small{}As shown in the proofs of Theorem 1 and 2, for any $S$,
ignoring $o_{P}(1\big/\sqrt{n}\Vert\beta_{X}\Vert_{2})$ and $o_{P}(\sqrt{p}\big/n\Vert\beta_{X}\Vert_{2}^{2})$
terms, 
\begin{eqnarray*}
\hat{\theta}_{C}-\theta_{0} & = & \frac{1}{\eta_{C}}\sum_{j\in S_{0}}J_{1j}\\
\hat{\theta}_{S_{k}}-\theta_{0}-\frac{b_{S_{k}}}{\eta_{C}+\eta_{S_{k}}} & = & \frac{1}{\eta_{C}+\eta_{S_{k}}}\sum_{j\in S_{0}\cup S_{k}}J_{1j}\\
\frac{\hat{b}_{S_{k}}-b_{S_{k}}}{\eta_{C}+\eta_{S_{k}}} & = & \frac{1}{\eta_{C}+\eta_{S_{k}}}\sum_{j\in S_{k}}\bar{B}_{j}-\frac{\eta_{S_{k}}}{\eta_{C}(\eta_{C}+\eta_{S_{k}})}\sum_{j\in S_{0}}J_{1j}
\end{eqnarray*}
where $J_{1j}=\Omega_{j}^{-1}\beta_{X_{j}}(e_{Y_{j}}-\theta_{0}e_{X_{j}})+\Omega_{j}^{-2}(e_{Y_{j}}-\theta_{0}e_{X_{j}})(\sigma_{Y_{j}}^{2}e_{X_{j}}+\theta_{0}\sigma_{X_{j}}^{2}e_{Y_{j}})$,
and $\bar{B}_{j}=\Omega_{j}^{-1}\beta_{X_{j}}(e_{Y_{j}}-\theta_{0}e_{X_{j}})+\Omega_{j}^{-1}e_{X_{j}}e_{Y_{j}}-\theta_{0}\Omega_{j}^{-1}(e_{X_{j}}^{2}-\sigma_{X_{j}}^{2}).$}{\small\par}

{\small{}We can partition the $K$ additional instrument sets into
$L\leq2^{K}-1$ }\textsl{\small{}distinct}{\small{} sets which span
the additional instrument sets $S_{1},...,S_{K}$. For example, for
$K=3$, each instrument must belong to one, and only one, of the following
sets: $M_{1}=S_{1}\cap S_{2}\cap S_{3}$, $M_{2}=S_{1}\cap S_{2}\cap S_{3}^{C}$,
$M_{3}=S_{1}\cap S_{2}^{C}\cap S_{3}$, $M_{4}=S_{1}^{C}\cap S_{2}\cap S_{3}$,
$M_{5}=S_{1}\cap S_{2}^{C}\cap S_{3}^{C}$, $M_{6}=S_{1}^{C}\cap S_{2}\cap S_{3}^{C}$,
and $M_{7}=S_{1}^{C}\cap S_{2}^{C}\cap S_{3}$. Then, for each $j\in[3]$,
we can construct selection indicators $\alpha_{\ell}\in\{0,1\}$,
$\ell\in[7]$, such that $S_{j}=\bigcup_{\ell=1}^{7}\alpha_{\ell}M_{\ell}$. }{\small\par}

{\small{}For $L\leq2^{\vert K\vert}-1$, let $M_{1},...,M_{L}$ be
distinct sets of the additional instruments which span the additional
instrument sets $S_{1},...,S_{K}$. }{\small\par}

{\small{}We can therefore write }{\small\par}

{\small{}
\[
\begin{pmatrix}\hat{\theta}_{C}-\theta_{0}\\
\hat{\theta}_{S_{1}}-\theta_{0}-\frac{b_{S_{1}}}{\eta_{C}+\eta_{S_{1}}}\\
\vdots\\
\hat{\theta}_{S_{K}}-\theta_{0}-\frac{b_{S_{K}}}{\eta_{C}+\eta_{S_{K}}}\\
\frac{\hat{b}_{S_{1}}-b_{S_{1}}}{\eta_{C}+\eta_{S_{1}}}\\
\vdots\\
\frac{\hat{b}_{S_{K}}-b_{S_{K}}}{\eta_{C}+\eta_{S_{K}}}
\end{pmatrix}=\pi_{C}\begin{pmatrix}\sum_{j\in S_{0}}J_{1j}\end{pmatrix}+\sum_{\ell=1}^{L}\pi_{M_{\ell}}\begin{pmatrix}\sum_{j\in M_{\ell}}\mu_{j}\end{pmatrix},
\]
where 
\[
\mu_{j}=\begin{pmatrix}J_{1j}\\
H_{j}
\end{pmatrix},\,\,\pi_{C}=\begin{pmatrix}\frac{1}{\eta_{C}}\\
\frac{1}{\eta_{C}+\eta_{S_{1}}}\\
\vdots\\
\frac{1}{\eta_{C}+\eta_{S_{K}}}\\
-\frac{\eta_{S_{1}}}{\eta_{C}(\eta_{C}+\eta_{S_{1}})}\\
\vdots\\
-\frac{\eta_{S_{K}}}{\eta_{C}(\eta_{C}+\eta_{S_{K}})}
\end{pmatrix},\,\,\pi_{M_{\ell}}=\begin{pmatrix}0 & 0\\
\frac{\mathbb{I}\{M_{\ell}\subseteq S_{1}\}}{\eta_{C}+\eta_{S_{1}}} & 0\\
\vdots & \vdots\\
\frac{\mathbb{I}\{M_{\ell}\subseteq S_{K}\}}{\eta_{C}+\eta_{S_{K}}} & 0\\
. & \frac{\mathbb{I}\{M_{\ell}\subseteq S_{1}\}}{\eta_{C}+\eta_{S_{1}}}\\
\vdots & \vdots\\
0 & \frac{\mathbb{I}\{M_{\ell}\subseteq S_{K}\}}{\eta_{C}+\eta_{S_{K}}}
\end{pmatrix}\begin{pmatrix}1 & 0\\
1 & -1
\end{pmatrix},\,(\ell=1,..,L).
\]
}{\small\par}

{\small{}and $H_{j}=\Omega_{j}^{-2}\theta_{0}(\sigma_{X_{j}}^{2}e_{Y_{j}}^{2}-\sigma_{Y_{j}}^{2}e_{X_{j}}^{2}-2\theta_{0}\sigma_{X_{j}}^{2}e_{X_{j}}e_{Y_{j}})+\theta_{0}\Omega_{j}^{-1}(e_{X_{j}}^{2}-\sigma_{X_{j}}^{2})$.
For any set $M_{\ell}$, we will show that 
\begin{equation}
\sum_{j\in M_{\ell}}\mu_{j}\overset{a}{\sim}N\left(\begin{bmatrix}0\\
0
\end{bmatrix},\begin{bmatrix}\eta_{M_{\ell}}+\varsigma_{M_{\ell}} & 0\\
0 & \xi_{M_{\ell}}
\end{bmatrix}\right).
\end{equation}
}{\small\par}

{\small{}Also, as in the Proof of Lemma S.3, we have 
\[
\sum_{j\in S_{0}}J_{1j}\overset{a}{\sim}N(0,\eta_{C}+\varsigma_{C}).
\]
Then, since (i) the random components in $J_{1j}$ and $\mu_{j}$
are functions of the error terms $e_{X_{j}}$ and $e_{Y_{j}}$; (ii)
for any $j\neq k$, $e_{X_{j}}$ and $e_{X_{k}}$ are jointly normal
and uncorrelated, and hence mutually independent (likewise for $e_{Y_{j}}$
and $e_{X_{j}}$), we have that $\sum_{j\in S_{0}}J_{1j}$, $\sum_{j\in M_{1}}\mu_{j}$,
..., $\sum_{j\in M_{K_{0}}}\mu_{j}$ are mutually independent sums.
Therefore, 
\[
\pi_{C}\begin{pmatrix}\sum_{j\in S_{0}}J_{1j}\end{pmatrix}+\sum_{\ell=1}^{L}\pi_{M_{\ell}}\begin{pmatrix}\sum_{j\in M_{\ell}}\mu_{j}\end{pmatrix}\overset{a}{\sim}N\left(\underset{((2K+1)\times1)}{0},R+W\right),
\]
where $R=\pi_{C}(\eta_{C}+\zeta_{C})\pi_{C}^{\prime}$ and $W=\sum_{\ell=1}^{L}\pi_{M_{\ell}}\begin{bmatrix}\eta_{M_{\ell}}+\varsigma_{M_{\ell}} & 0\\
0 & \xi_{M_{\ell}}
\end{bmatrix}\pi_{M_{\ell}}^{\prime}$. }{\small\par}

{\small{}Some straight-forward calculations show that 
\[
R=\begin{bmatrix}\underset{(1\times1)}{R_{11}} & \underset{(1\times K)}{R_{21}^{\prime}} & \underset{(1\times K)}{R_{31}^{\prime}}\\
\underset{(K\times1)}{R_{21}} & \underset{(K\times K)}{R_{22}} & \underset{(K\times K)}{R_{32}^{\prime}}\\
\underset{(K\times1)}{R_{31}} & \underset{(K\times K)}{R_{32}} & \underset{(K\times K)}{R_{33}}
\end{bmatrix},
\]
where $R_{11}=\eta_{C}^{-2}(\eta_{C}+\varsigma_{C})$, the $k$-th
element of $R_{21}$ is given by $R_{21}^{(k)}=\eta_{C}^{-1}(\eta_{C}+\eta_{S_{k}})^{-1}(\eta_{C}+\varsigma_{C})$,
the $k$-th element of $R_{31}$ is given by $R_{31}^{(k)}=-\eta_{C}^{-2}(\eta_{C}+\eta_{S_{k}})^{-1}\eta_{S_{k}}(\eta_{C}+\varsigma_{C})$,
the $(k,l)$-th element of $R_{22}$ is given by $R_{22}^{(k,l)}=(\eta_{C}+\eta_{S_{k}})^{-1}(\eta_{C}+\eta_{S_{l}})^{-1}(\eta_{C}+\varsigma_{C})$,
the $(k,l)$-th element of $R_{32}$ is given by $R_{32}^{(k,l)}=-\eta_{C}^{-1}\eta_{S_{k}}(\eta_{C}+\eta_{S_{k}})^{-1}(\eta_{C}+\eta_{S_{l}})^{-1}(\eta_{C}+\varsigma_{C})$,
and the $(k,l)$-th element of $R_{33}$ is given by $R_{33}^{(k,l)}=\eta_{C}^{-2}(\eta_{C}+\eta_{S_{k}})^{-1}(\eta_{C}+\eta_{S_{l}})^{-1}\eta_{S_{k}}\eta_{S_{k}}(\eta_{C}+\varsigma_{C})$,
$(k,l=1,...,K)$. }{\small\par}

{\small{}Similarly, we have 
\[
W=\begin{bmatrix}\underset{(1\times1)}{0} & \underset{(1\times K)}{0} & \underset{(1\times K)}{0}\\
\underset{(K\times1)}{0} & \underset{(K\times K)}{W_{1}} & \underset{(K\times K)}{W_{1}}\\
\underset{(K\times1)}{0} & \underset{(K\times K)}{W_{1}} & \underset{(K\times K)}{W_{2}}
\end{bmatrix},
\]
where the $(k,l)$-th element of $W_{1}$ is given by $W_{1}^{(k,l)}=(\eta_{C}+\eta_{S_{k}})^{-1}(\eta_{C}+\eta_{S_{l}})^{-1}(\eta_{S_{k}\cap S_{l}}+\varsigma_{S_{k}\cap S_{l}})$,
and the $(k,l)$-th element of $W_{2}$ is given by $W_{2}^{(k,l)}=(\eta_{C}+\eta_{S_{k}})^{-1}(\eta_{C}+\eta_{S_{l}})^{-1}(\eta_{S_{k}\cap S_{l}}+\varsigma_{S_{k}\cap S_{l}}+\xi_{S_{k}\cap S_{l}})$,
$(k,l=1,...,K)$.}{\small\par}

{\small{}The expression for the covariance matrix in Theorem 3 is
then $\Delta=R+W$.}{\small\par}

\textbf{\small{}Proof of Equation 4.}{\small\par}

{\small{}If $\vert M_{\ell}\vert=o(p)$, then the following asymptotic
distribution result for $\sum_{j\in M_{\ell}}\mu_{j}$ still applies
but variance components $\varsigma_{M_{\ell}}$ and $\xi_{M_{\ell}}$
from Equation 4 would be negligible. Therefore, we focus on the case
$\vert M_{\ell}\vert=O(p)$ for $\ell\in[L]$. We wish to apply the
following multivariate Berry-Esseen result by \citet{Bentkus2005}.
For a subset of instruments $M_{\ell}$ and the $2$-dimensional vector
of independent variables $\mu_{1},...,\mu_{\vert M_{\ell}\vert}$,
let $\Delta_{\ell}=Var\big(\sum_{j\in M_{\ell}}\mu_{j}\big)$. Then
for $U\sim N(0,\Delta_{\ell})$, for any convex set $S$, 
\[
\big\vert P\big(\sum_{j\in M_{\ell}}\mu_{j}\in S\big)-P(U\in S)\big\vert\leq O\Big(\sum_{j\in M_{\ell}}E[\Vert\Delta_{\ell}^{-\frac{1}{2}}\mu_{j}\Vert_{2}^{3}]\Big).
\]
}{\small\par}

{\small{}Thus, joint convergence of effect and bias estimates follows
by the following equations
\begin{align}
\Delta_{\ell} & =\Theta(n\Vert\beta_{X}\Vert_{2}^{2}+p)\\
\sum_{j\in M_{\ell}}E[\Vert\mu_{j}\Vert_{2}^{3}] & =O(n^{\frac{3}{2}}\Vert\beta_{X}\Vert_{4}^{2}\Vert\beta_{X}\Vert_{2})+O(\sqrt{p}n\Vert\beta_{X}\Vert_{4}^{2})+O(n\Vert\beta_{X}\Vert_{2}^{2})+O(p)
\end{align}
since $(5)$ implies $\Delta_{\ell}^{-\frac{1}{2}}=\Theta(1\big/\sqrt{n}\Vert\beta_{X}\Vert_{2})+\Theta(1\big/\sqrt{p})$,
and by $(6)$, 
\begin{eqnarray*}
\sum_{j\in M_{\ell}}E[\Vert\Delta_{\ell}^{-\frac{1}{2}}\mu_{j}\Vert_{2}^{3}] & = & \Delta_{\ell}^{-\frac{3}{2}}\sum_{j\in M_{\ell}}E[\Vert\mu_{j}\Vert_{2}^{3}]\\
 & = & O\Big(\frac{\Vert\beta_{X}\Vert_{4}^{2}}{\Vert\beta_{X}\Vert_{2}^{2}}\Big)+O\Big(\frac{\sqrt{p}}{\sqrt{n}\Vert\beta_{X}\Vert_{2}}\Big)O\Big(\frac{\Vert\beta_{X}\Vert_{4}^{2}}{\Vert\beta_{X}\Vert_{2}^{2}}\Big)+O\Big(\frac{1}{\sqrt{n}\Vert\beta_{X}\Vert_{2}}\Big)+O\Big(\frac{1}{\sqrt{p}}\Big)\\
 & = & o(1),
\end{eqnarray*}
as $\Vert\beta_{X}\Vert_{4}\big/\Vert\beta_{X}\Vert_{2}\to0$ and
$p\big/n\Vert\beta_{X}\Vert_{2}^{2}=O(1)$ by Assumption 3. }{\small\par}

{\small{}To show $(5)$, we calculate the covariance matrix $\Delta_{\ell}=Var(\sum_{j\in M_{\ell}}\mu_{j})$
as 
\begin{eqnarray*}
\Delta_{\ell} & = & \sum_{j\in M_{\ell}}Var(\mu_{j})\\
 & = & \begin{bmatrix}\eta_{M_{\ell}}+\varsigma_{M_{\ell}} & 0\\
0 & \xi_{M_{\ell}}
\end{bmatrix}\\
 & \,\\
 & = & \Theta(n\Vert\beta_{X}\Vert_{2}^{2}+p),
\end{eqnarray*}
since $Var(J_{1j})=\Omega_{j}^{-1}\beta_{X_{j}}^{2}+\Omega_{j}^{-2}\sigma_{X_{j}}^{2}\sigma_{Y_{j}}^{2}$,
$Cov(J_{1j},H_{j})=0$, and $Var(H_{j})=2\theta_{0}^{2}\Omega_{j}^{-2}\sigma_{X_{j}}^{4}$. }{\small\par}

{\small{}To show $(6)$, note that $\Vert\mu_{j}\Vert_{2}^{2}=J_{1j}^{2}+(J_{1j}-\bar{B}_{j})^{2}\leq3J_{1j}^{2}+2\bar{B}_{j}^{2}$,
and $\Vert\mu_{j}\Vert_{2}^{4}\leq9J_{1j}^{4}+12J_{1j}^{2}\bar{B}_{j}^{2}+4\bar{B}_{j}^{4}$.
Thus, $E[\Vert\mu_{j}\Vert_{2}^{2}]=3E[J_{1j}^{2}]+2E[\bar{B}_{j}^{2}]$,
and $E[\Vert\mu_{j}\Vert_{2}^{4}]\leq15E[J_{1j}^{4}]+10E[\bar{B}_{1j}^{4}]$. }{\small\par}

{\small{}Then, 
\begin{eqnarray*}
\sum_{j\in M_{\ell}}E[\Vert\mu_{j}\Vert_{2}^{2}] & = & 3\sum_{j\in M_{\ell}}E[J_{1j}^{2}]+2\sum_{j\in M_{\ell}}E[\bar{B}_{j}^{2}]\\
 & = & 5\eta_{M_{\ell}}+5\varsigma_{M_{\ell}}+2\xi_{M_{\ell}}\\
 & = & \Theta(n\Vert\beta_{X}\Vert_{2}^{2}+p),
\end{eqnarray*}
and 
\begin{eqnarray*}
\sum_{j\in M_{\ell}}E[\Vert\mu_{j}\Vert_{2}^{4}] & \leq & 15\sum_{j\in M_{\ell}}E[J_{1j}^{4}]+10\sum_{j\in M_{\ell}}E[\bar{B}_{j}^{4}]\\
 & = & O(n^{2}\Vert\beta_{X}\Vert_{4}^{4})+O(n\Vert\beta_{X}\Vert_{2}^{2})+O(p),
\end{eqnarray*}
since by direct calculation, it can be shown that there exist constants
$C_{l}$, $l\in[10]$ such that 
\begin{eqnarray*}
\sum_{j\in M_{\ell}}E[J_{1j}^{4}] & = & C_{1}\sum_{j\in M_{\ell}}\Omega_{j}^{-2}\beta_{X_{j}}^{4}+C_{2}\sum_{j\in M_{\ell}}\Omega_{j}^{-3}\sigma_{X_{j}}^{2}\sigma_{Y_{j}}^{2}\beta_{X_{j}}^{2}+C_{3}\sum_{j\in M_{\ell}}\Omega_{j}^{-4}\sigma_{X_{j}}^{4}\sigma_{X_{j}}^{4}\\
 & = & \Theta(n^{2}\Vert\beta_{X}\Vert_{4}^{4})+\Theta(n\Vert\beta_{X}\Vert_{2}^{2})+\Theta(p),
\end{eqnarray*}
and 
\begin{eqnarray*}
\sum_{j\in M_{\ell}}E[\bar{B}_{j}^{4}] & = & C_{4}\sum_{j\in M_{\ell}}\Omega_{j}^{-2}\beta_{X_{j}}^{4}+C_{5}\sum_{j\in M_{\ell}}\Omega_{j}^{-2}\sigma_{X_{j}}^{2}\beta_{X_{j}}^{2}+C_{6}\theta_{0}^{2}\sum_{j\in M_{\ell}}\Omega_{j}^{-3}\sigma_{X_{j}}^{4}\beta_{X_{j}}^{2}+C_{7}\theta_{0}^{2}\sum_{j\in M_{\ell}}\Omega_{j}^{-4}\sigma_{X_{j}}^{4}\sigma_{Y_{j}}^{2}\beta_{X_{j}}^{2}\\
 &  & +C_{8}\sum_{j\in M_{\ell}}\Omega_{j}^{-2}\sigma_{X_{j}}^{4}+C_{9}\theta_{0}^{2}\sum_{j\in M_{\ell}}\Omega_{j}^{-3}\sigma_{X_{j}}^{6}+C_{10}\theta_{0}^{4}\sum_{j\in M_{\ell}}\Omega_{j}^{-4}\sigma_{X_{j}}^{4}\sigma_{Y_{j}}^{4}\\
 & = & \Theta(n^{2}\Vert\beta_{X}\Vert_{4}^{4})+\Theta(n\Vert\beta_{X}\Vert_{2}^{2})+\Theta(p).
\end{eqnarray*}
Thus, by CS and the above, 
\begin{eqnarray*}
\sum_{j\in M_{\ell}}E[\Vert\mu_{j}\Vert_{2}^{3}] & \leq & \Big(\sum_{j\in M_{\ell}}E[\Vert\mu_{j}\Vert_{2}^{4}]\Big)^{\frac{1}{2}}\Big(\sum_{j\in M_{\ell}}E[\Vert\mu_{j}\Vert_{2}^{2}]\Big)^{\frac{1}{2}}\\
 & \leq & \big(O(n^{2}\Vert\beta_{X}\Vert_{4}^{4})+O(n\Vert\beta_{X}\Vert_{2}^{2})+O(p)\big)^{\frac{1}{2}}\big(O(n\Vert\beta_{X}\Vert_{2}^{2})+O(p)\big)^{\frac{1}{2}}\\
 & = & O(n^{\frac{3}{2}}\Vert\beta_{X}\Vert_{4}^{2}\Vert\beta_{X}\Vert_{2})+O(\sqrt{p}n\Vert\beta_{X}\Vert_{4}^{2})+O(n\Vert\beta_{X}\Vert_{2}^{2})+O(p),
\end{eqnarray*}
as required.\hfill{}$\square$}{\small\par}

~

\textbf{Proof of Theorem 3 -- Part II (Asymptotic distribution of
the Focused estimator $\hat{\theta}$).}

{\small{}Let $\hat{\omega}_{C}$ denote a binary indicator that equals
1 only when $S_{0}$ is the AMSE minimising set of instruments, and
let $\hat{\omega}_{k}$, $k\in[K]$ denote binary indicators which
which equal 1 only when $S_{0}\cup S_{k}$ is the AMSE minimising
set of instruments. Then, $\hat{\omega}_{C}+\sum_{k=1}^{K}\hat{\omega}_{k}=1$,
and the Focused estimator can be written 
\[
\hat{\theta}-\theta_{0}=\hat{\omega}_{C}(\hat{\theta}_{S_{0}}-\theta_{0})+\sum_{k=1}^{K}\hat{\omega}_{k}(\hat{\theta}_{S_{k}}-\theta_{0}).
\]
Let $U=(U_{1},\ldots,U_{2K+1})^{\prime}$ be the normally distributed
vector $U\sim N(0,\Delta)$. Then, under Assumptions 1-4, the asymptotic
distribution of the Focused estimator is 
\[
\hat{\theta}-\theta_{0}\overset{a}{\sim}\omega_{C}^{\star}U_{1}+\sum_{k=1}^{K}\omega_{k}^{\star}\Big[U_{k+1}+\frac{b_{S_{k}}}{\eta_{C}+\eta_{S_{k}}}\Big],
\]
as $n,p\to\infty$, where $\omega_{C}^{\star}=I\big\{\Delta_{C}\leq\min_{k^{\prime}\in[K]}\big([U_{K+k^{\prime}+1}+(\eta_{C}+\eta_{S_{k^{\prime}}})^{-2}b_{S_{k^{\prime}}}]^{2}-\Delta_{B}^{(k^{\prime},k^{\prime})}+\Delta_{F}^{(k^{\prime},k^{\prime})}\big)\big\}$
and $\omega_{k}^{\star}=(1-\omega_{C}^{\star})\times I\big\{[U_{K+k+1}+(\eta_{C}+\eta_{S_{k}})^{-2}b_{S_{k}}]^{2}-\Delta_{B}^{(k,k)}+\Delta_{F}^{(k,k)}=\min_{k^{\prime}\in[K]}\big([U_{K+k^{\prime}+1}+(\eta_{C}+\eta_{S_{k^{\prime}}})^{-2}b_{S_{k^{\prime}}}]^{2}-\Delta_{B}^{(k^{\prime},k^{\prime})}+\Delta_{F}^{(k^{\prime},k^{\prime})}\big)\big\}$,
$k\in[K]$. }{\small\par}

{\small{}To see this, note that for all $k\in[K]$, $\hat{\omega}_{C}$
and $\hat{\omega}_{k}$ are functions of the estimated AMSE which
consist of consistent estimators of constants (see Lemma S.5) and
$\hat{b}_{S_{k}}$. The result then follows by Slutsky's lemma and
Part I which shows the joint convergence in distribution of $\hat{\theta}_{C}$,
$\hat{b}_{S_{k}}$ and $\hat{\theta}_{S_{k}}$ over all instruments
sets $S_{0}\cup S_{k}$, $k\in[K]$. \hfill{}$\square$}{\small\par}

~

\textbf{Proof of Theorem 4 (Worst case size distortion of Focused
intervals).}

{\small{}Let $\hat{b}=\big(\hat{b}_{S_{1}}\big/(\hat{\eta}_{C}+\hat{\eta}_{S_{1}}),\ldots,\hat{b}_{S_{K}}\big/(\hat{\eta}_{C}+\hat{\eta}_{S_{K}})\big)^{\prime}$
and $b=\big(b_{S_{1}}\big/(\eta_{C}+\eta_{S_{1}}),\ldots,b_{S_{K}}\big/(\eta_{C}+\eta_{S_{K}})\big)^{\prime}$.
Let ${\cal B}(b,\alpha_{1})=\{b^{\star}:\,\Gamma(b,b^{\star})\leq\chi_{K}^{2}(\alpha_{1})\}$
where $\chi_{K}^{2}(\alpha_{1})$ denotes the $1-\alpha_{1}$ quantile
of a random $\chi_{K}^{2}$ variable, $\Gamma(b,b^{\star})=(b+M-b^{\star})^{\prime}\Delta_{B}^{-1}(b+M-b^{\star})$,
and $M\sim N(0_{K\times1},\Delta_{B})$. Therefore, by Theorem 2,
${\cal B}(b,\alpha_{1})$ is the limiting version of a $(1-\alpha_{1})\times100\%$
confidence region for the true asymptotic bias $b$. }{\small\par}

{\small{}Let ${\cal C}=(a_{L}(b'),a_{U}(b'))$ define a collection
of $(1-\alpha_{2})\times100\%$ confidence intervals indexed by $b'$,
each constructed such that 
\begin{eqnarray*}
P\big(a_{L}(b')\leq\Lambda(b')\leq a_{U}(b')\big)=1-\alpha_{2} & \text{if} & b'=b,\,\,\,\,\,\text{and}\\
P\big(a_{L}(b')\leq\Lambda(b'')\leq a_{U}(b')\big)\geq1-\alpha_{2}-\gamma & \text{for all} & b''\in{\cal B}(b,\alpha_{1}).
\end{eqnarray*}
Let $(a_{L}^{\star},a_{U}^{\star})=(a_{L}(b^{\star}),a_{U}(b^{\star}))$
be an interval such that $a_{U}(b^{\star})-a_{L}(b^{\star})\leq a_{U}(b')-a_{L}(b')$
for all intervals $(a_{L}(b^{\star}),a_{U}(b^{\star}))$ and $(a_{L}(b'),a_{U}(b'))$
contained in ${\cal C}$, and $(b^{\star},b')\in{\cal B}(b,\alpha_{1})$. }{\small\par}

{\small{}Let $A=\{\Gamma(b,b)\leq\chi_{K}^{2}(\alpha_{1})\}$, so
that $A$ is the event that the limiting version of the confidence
region for the asymptotic bias contains the true asymptotic bias $b$.
Note that since $\Gamma(b,b^{\star})\sim\chi_{K}^{2}$, we have $P(A)=1-\alpha_{1}$. }{\small\par}

{\small{}For every $b'\in{\cal B}(b,\alpha_{1})$, we have $P\big(\{a_{L}^{\star}\leq\Lambda(b')\leq a_{U}^{\star}\}\cap A\big)+P\big(\{a_{L}^{\star}\leq\Lambda(b')\leq a_{U}^{\star}\}\cap A^{c}\big)\geq1-\alpha_{2}-\gamma$
since $(a_{L}^{\star},a_{U}^{\star})\in{\cal C}$. Moreover, note
that $P\big(\{a_{L}^{\star}\leq\Lambda(b')\leq a_{U}^{\star}\}\cap A^{c}\big)\leq P(A^{c})=\alpha_{1}$,
and therefore, $P\big(\{a_{L}^{\star}\leq\Lambda(b')\leq a_{U}^{\star}\}\cap A\big)\geq1-\alpha_{2}-\gamma-\alpha_{1}$
for all $b'\in{\cal B}(b,\alpha_{1})$. }{\small\par}

{\small{}If $A$ occurs, then by definition $b\in{\cal B}(b,\alpha_{1})$.
Hence, $P(\{a_{L}^{\star}\leq\Lambda(b)\leq a_{U}^{\star}\})\geq1-\alpha_{2}-\gamma-\alpha_{1}$.
We can repeat these steps for different combinations of $\alpha_{1}$
and $\alpha_{2}$ such that $\alpha_{1}+\alpha_{2}=\alpha$ which
leads to the result of the theorem. \hfill{}$\square$}{\small\par}

\section{Additional simulation results}

{\small{}For the design described in Section 5.1, Figures S1--S3
show the results of the Focused estimator and interval for weaker
instruments (the concentration parameter ranges from 10 to 50, compared
with 40 to 200 in the main text). Figure S4 shows the results of confidence
intervals when $S_{0}$ consists of invalid instruments, according
to the design discussed in Section 5.5. We note that the patterns
observed in Figures S1--S3 are very similar to those observed in
Figures 2--4 in the main text, highlighting that for finite-sample
performance, the relative strengths of the valid and additional instrument
sets may be more important than their absolute strengths. }{\small\par}
\begin{center}
\includegraphics[width=16.5cm]{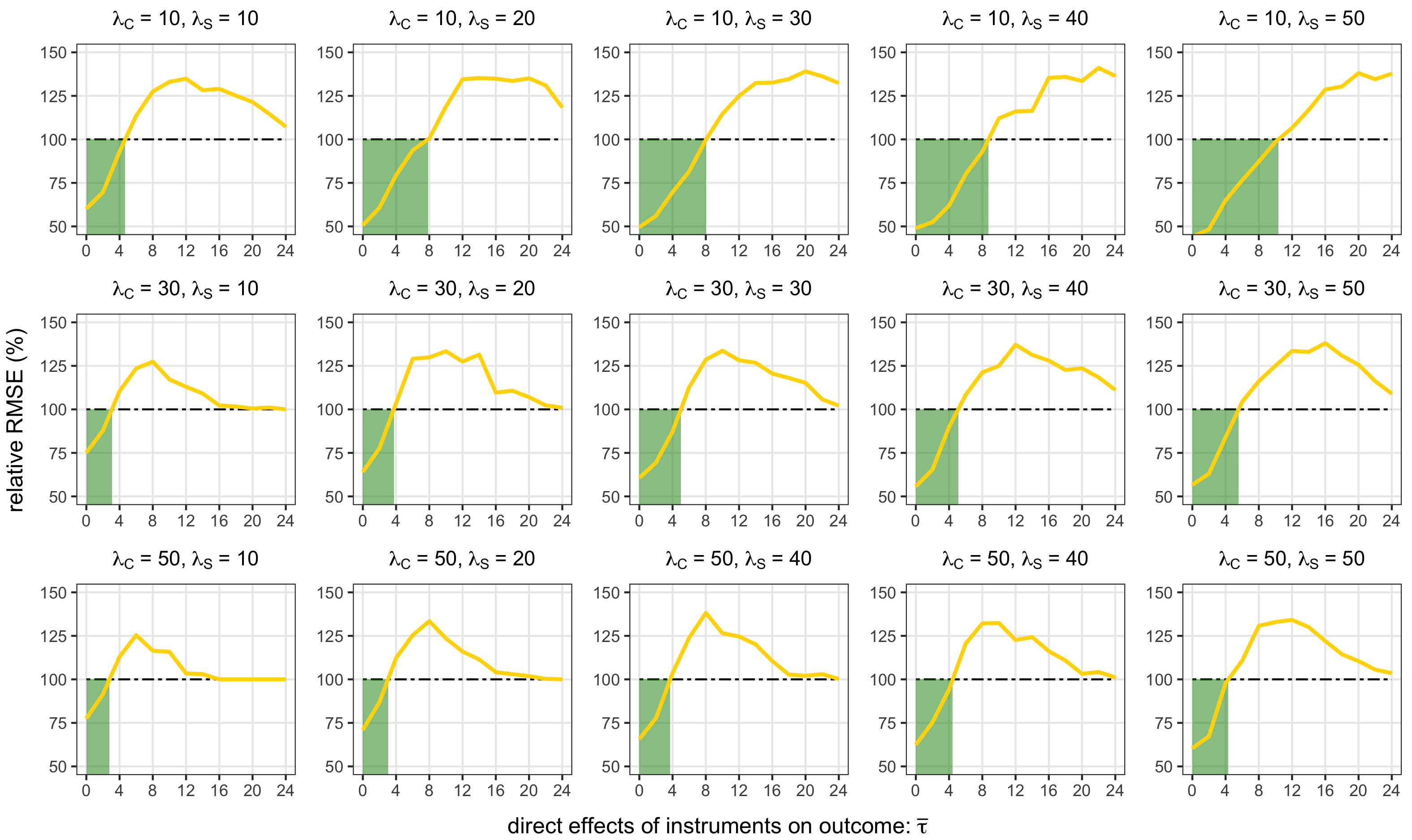}\\
{\small{}Figure S1. }{\footnotesize{}RMSE varying with the average
instrument strength of $S_{0}$ $(\lambda_{C})$ and $S$ $(\lambda_{S})$,
and invalidness of $S$ $(\bar{\tau})$. }{\footnotesize\par}
\par\end{center}

\begin{center}
\includegraphics[width=16.5cm]{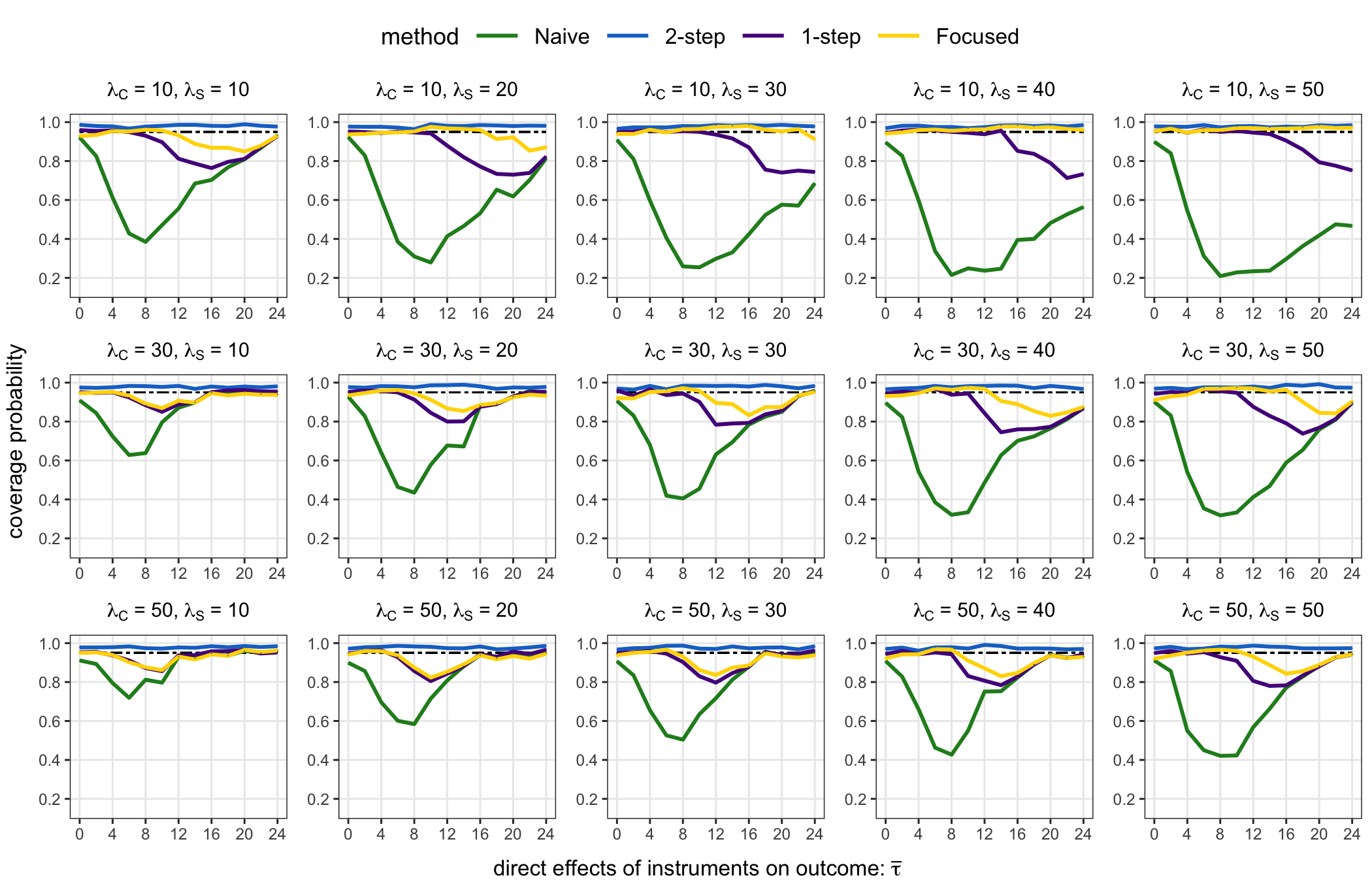}\\
{\small{}Figure S2. }{\footnotesize{}Coverage probabilities of confidence
intervals (nominal coverage is $1-\alpha=0.95$). }{\footnotesize\par}
\par\end{center}

\begin{center}
\includegraphics[width=16.5cm]{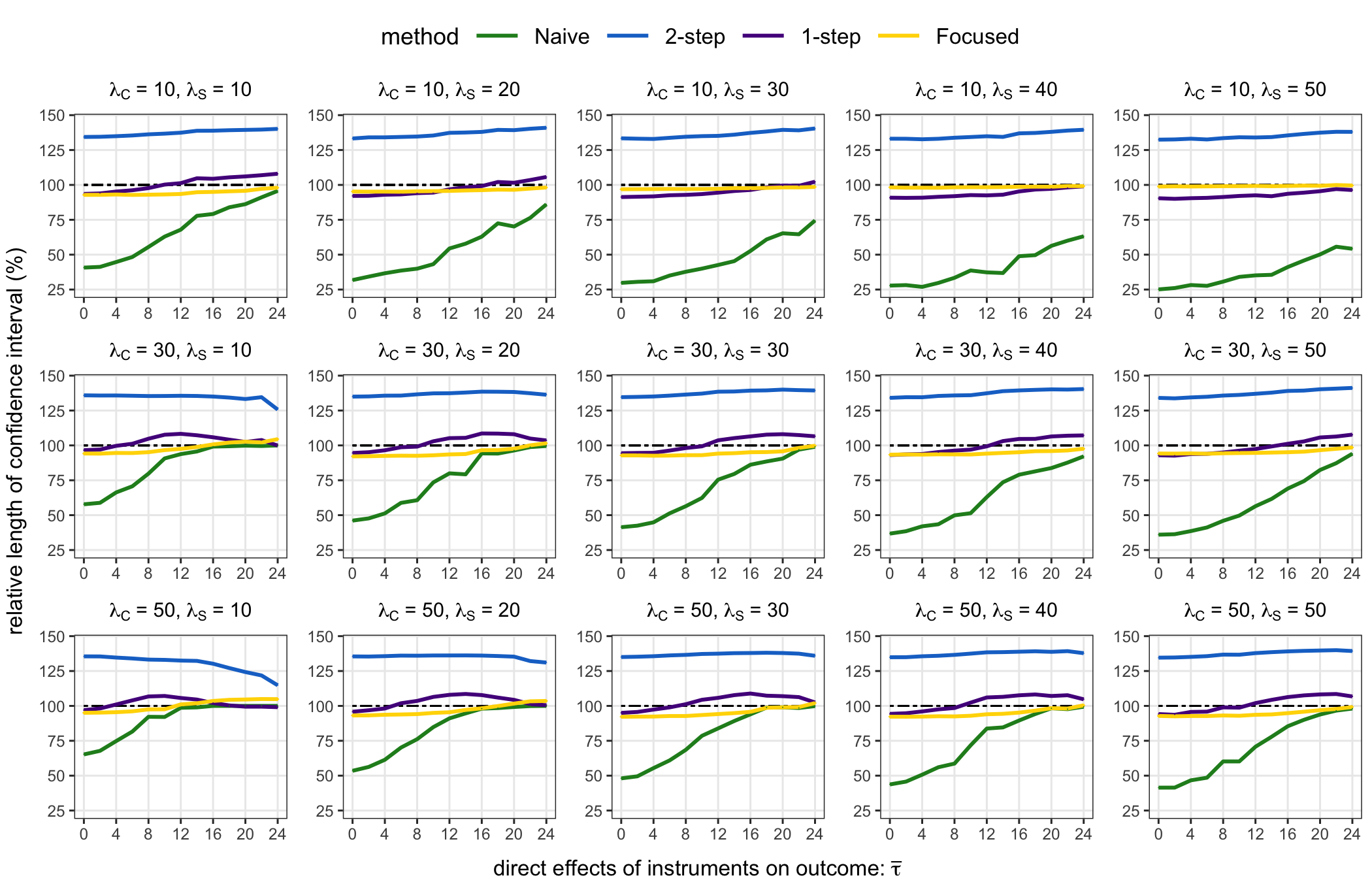}\\
{\small{}Figure S3. }{\footnotesize{}Length of confidence intervals
relative to the Core interval (nominal coverage is $1-\alpha=0.95$).
}\\
{\footnotesize{}~}\\
{\footnotesize{}~}{\footnotesize\par}
\par\end{center}

\begin{center}
\includegraphics[width=16.5cm]{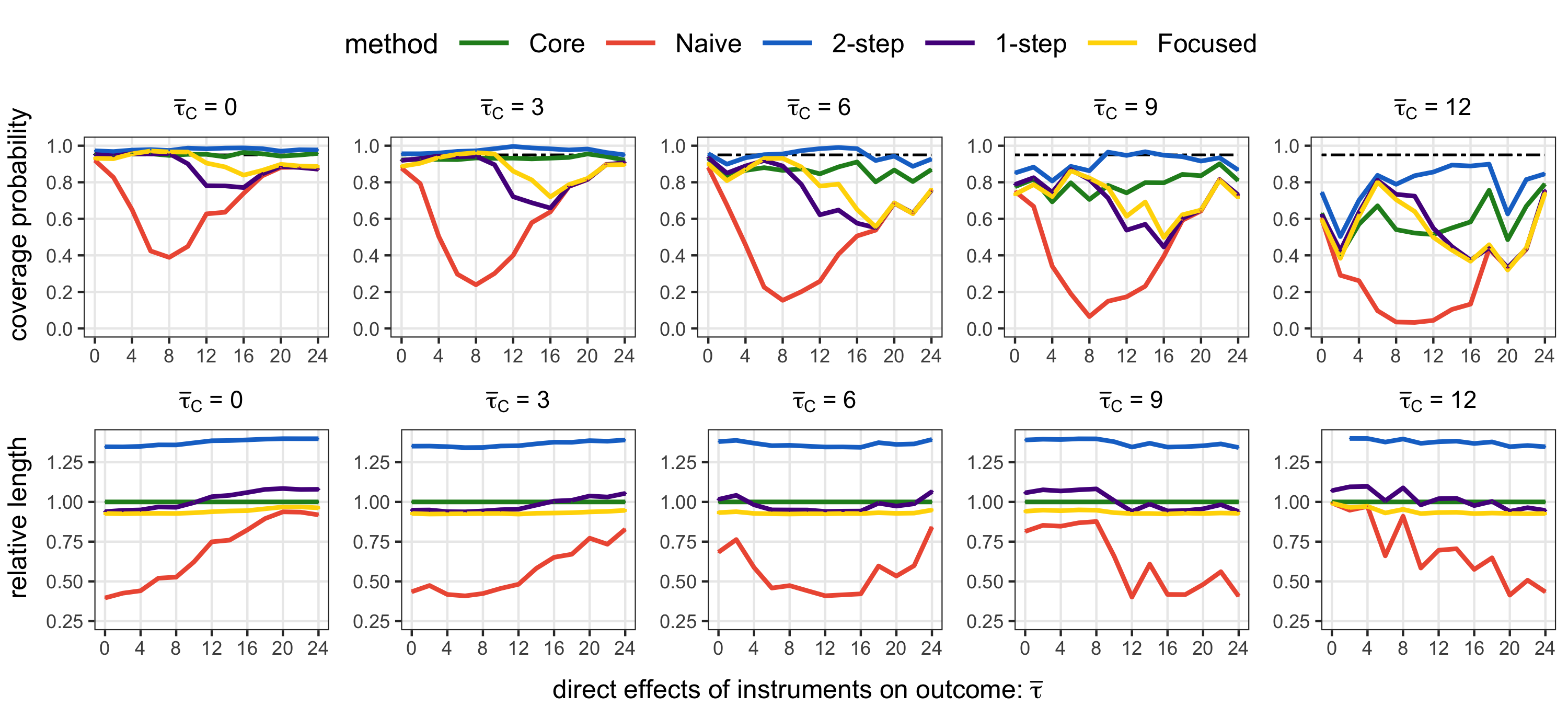}\\
{\small{}Figure S4. }{\footnotesize{}Coverage and length of 95\% confidence
intervals varying with invalidness of $S_{0}$ ($\bar{\tau}_{C})$
and invalidness of $S$ $(\bar{\tau})$. }{\footnotesize\par}
\par\end{center}

\newpage{}

{\small{}\bibliographystyle{chicago}
\bibliography{focusedMR}
}{\small\par}
\end{document}